\documentclass[twocolumn]{aastex62}
\usepackage{natbib}
\usepackage{amsmath}
\usepackage{gensymb}

\newcommand{\kms}         {km\,s$^{-1}$} 
\newcommand{\masyr}        {mas\,yr$^{-1}$}

\shorttitle{The Origin of the 300~\kms\ Stream}
\shortauthors{Fu et al.}

\begin{document}

\title{The Origin of the 300~km~s$^{-1}$ Stream Near Segue~1}

\author{Sal Wanying Fu}
\affiliation{Department of Physics and Astronomy, Pomona College, Claremont, CA 91711}

\author{Joshua D. Simon}
\affiliation{Observatories of the Carnegie Institution for Science, 813 Santa Barbara Street, Pasadena, CA 91101, USA}

\author{Matthew Shetrone}
\affiliation{University of Texas at Austin, McDonald Observatory, Fort Davis, TX 79734, USA}

\author{Jo Bovy}
\altaffiliation{Alfred P. Sloan Fellow}
\affiliation{Department of Astronomy and Astrophysics, University of Toronto, 50 St. George Street, Toronto, ON M5S 3H4, Canada}

\author{Timothy C. Beers}
\affiliation{Department of Physics and JINA Center for the Evolution of the Elements, University of Notre Dame, Notre Dame, IN 46556, USA}

\author{J. G. Fern{\'a}ndez-Trincado}
\affiliation{Departamento de Astronom\'\i a, Casilla 160-C, Universidad de Concepci\'on, Concepci\'on, Chile}
\affiliation{Instituto de Astronom\'ia y Ciencias Planetarias, Universidad de Atacama, Copayapu 485, Copiap\'o, Chile}
\affiliation{Institut Utinam, CNRS UMR 6213, Universit\'e Bourgogne-Franche-Comt\'e, OSU THETA Franche-Comt\'e, Observatoire de Besan\c{c}on, \\ BP 1615, 25010 Besan\c{c}on Cedex, France}

\author{Vinicius M. Placco}
\affiliation{Department of Physics and JINA Center for the Evolution of the Elements, University of Notre Dame, Notre Dame, IN 46556, USA}

\author{Olga Zamora}
\affiliation{Instituto de Astrof{\'i}sica de Canarias, E-38205 La Laguna, Tenerife, Spain}
\affiliation{Universidad de La Laguna (ULL), Departamento de Astrof{\'i}sica, E-38206 La Laguna, Tenerife, Spain}

\author{Carlos Allende Prieto}
\affiliation{Instituto de Astrof{\'i}sica de Canarias, E-38205 La Laguna, Tenerife, Spain}
\affiliation{Universidad de La Laguna (ULL), Departamento de Astrof{\'i}sica, E-38206 La Laguna, Tenerife, Spain}

\author{D. A. Garc{\'i}a-Hern{\'a}ndez}
\affiliation{Instituto de Astrof{\'i}sica de Canarias, E-38205 La Laguna, Tenerife, Spain}
\affiliation{Universidad de La Laguna (ULL), Departamento de Astrof{\'i}sica, E-38206 La Laguna, Tenerife, Spain}

\author{Paul Harding}
\affiliation{Department of Astronomy, Case Western Reserve University, Cleveland, OH 44106, USA}

\author{Inese Ivans}
\affiliation{Department of Physics \& Astronomy, University of Utah, Salt Lake City, UT, 84112, USA}

\author{Richard Lane}
\affiliation{Instituto de Astrof{\'i}sica at Pontificia Universidad Cat{\'o}lica de Chile, Av. Vicu\~na Mackenna 4860, 782-0436 Macul, Santiago, Chile}

\author{Christian Nitschelm}
\affiliation{Unidad de Astronom{\'i}a, Universidad de Antofagasta, Avenida Angamos 601, Antofagasta 1270300, Chile}

\author{Alexandre Roman-Lopes}
\affiliation{Department of Physics and Astronomy, Universidad de La Serena, Av. Juan Cisternas, 1200 North, La Serena, Chile}

\author{Jennifer Sobeck}
\affiliation{Department of Astronomy, University of Washington, Box 351580, Seattle, WA 98195, USA}

\correspondingauthor{Sal Wanying Fu}
\email{wanying.fu@pomona.edu}

\begin{abstract}

We present a search for new members of the 300~\kms\ stream (300S) near the dwarf galaxy Segue 1 using wide-field survey data. We identify 11 previously unknown bright stream members in the APOGEE-2 and SEGUE-1 and 2 spectroscopic surveys.  Based on the spatial distribution of the high-velocity stars, we confirm for the first time that this kinematic structure is associated with a 24\degr-long stream seen in SDSS and Pan-STARRS imaging data. The 300S stars display a metallicity range of $-2.17 < {\rm [Fe/H]} < -1.24$, with an intrinsic dispersion of 0.21$_{-0.09}^{+0.12}$~dex. They also have chemical abundance patterns similar to those of Local Group dwarf galaxies, as well as that of the Milky Way halo. Using the open-source code \textit{galpy} to model the orbit of the stream, we find that the progenitor of the stream passed perigalacticon about 70~Myr ago, with a closest approach to the Galactic Center of about 4.1~kpc. Using Pan-STARRS DR1 data, we obtain an integrated stream luminosity of $4 \times 10^3$~L$_{\odot}$. We conclude that the progenitor of the stream was a dwarf galaxy that is probably similar to the satellites that were accreted to build the present-day Milky Way halo. 
\end{abstract}



\section{Introduction}

\par The $\Lambda$CDM cosmological model posits that larger galaxies form via the hierarchical mergers of smaller galaxies \citep[e.g.,][]{searle1978compositions,white1978}. Simulations using $\Lambda$CDM predict that hierarchical merging processes form stellar substructures, remnants of individually tidally disrupted dwarf satellite galaxies, in the halos of larger galaxies \citep[e.g.,][]{bullock2001hierarchical,bullock2005,cooper2010}. The Milky Way halo, which we can study in great detail, encodes the history of tidal accretion events in the form of stellar streams and other substructures. Studying these streams can provide clues to the Milky Way's recent formation history, such as the timescale of accretion events, as well as the type of objects that were accreted to form the Galaxy's halo \citep[e.g.,][]{helmi1999,zolotov2010,bonaca2012cold,tissera2014a,koposov2014discovery,pillepich2015,guglielmo2017origin,kupper2017exploding}.

\par Multiple spectroscopic studies of the dwarf galaxy Segue~1 have uncovered a distinct population of stars with a heliocentric recessional velocity of $\sim300$~\kms\ in the same field (\citealt[][henceforth G09]{geha2009least}, \citealt[][henceforth N10]{norris2010chemical}, \citealt[][henceforth S11]{simon2011complete}).  S11 suggested that these stars belong to a stellar stream (henceforth 300S) rather than a compact object due to the stars' diffuse positions over their $\sim0.25\degr$ survey area. \citet[][henceforth F13]{frebel2013300} studied the chemical abundances of the brightest star in the stellar stream and suggested that it is similar to halo stars; this means that the stream's only distinct signature from the halo is its 300~\kms\ heliocentric velocity. If that is the case, then 300S may be representative of a large number of similar objects that were accreted at earlier times. However, the nature of the progenitor of 300S is currently unknown. 

\par From SDSS photometric data, \cite{niederste2009origin} used a matched-filter technique to discover an elongated structure in the vicinity of Segue~1, extending about 4$\degree$ east-west. S11 suggested that this feature could be the photometric counterpart of 300S. \citet[][henceforth B16]{bernard2016synoptic} used Pan-STARRS photometry to trace the same feature over a wider area of the sky, showing that it extends spatially over the range $144\degr < \mbox{RA} < 168\degr$ (also see \citealt{grillmair2014}). Follow-up spectroscopic observations over this patch of the sky are crucial for determining whether the photometric structure and the kinematic structure are linked.

\par All-sky surveys are optimal for studying stellar streams because of (1) their ability to detect and map out the full extent of the stream, and (2) their ability to provide photometric and spectroscopic data that allows for further characterization of stellar streams. In particular, radial velocities and proper motions allow us to model the orbit of the stream, providing insight into the stream's tidal disruption history. All-sky, high-resolution spectroscopic surveys such as APOGEE-2 can also yield detailed insights into the chemical evolution of the stream progenitor, and discern whether the progenitor is a globular cluster or a dwarf galaxy. 

\par In this study, we present an analysis of members of 300S found in APOGEE-2 and SEGUE data. Section \ref{sec:new_members} describes the member selection process. In Section \ref{sec:abundances}, we compare the chemical abundances of 300S to those of other Milky Way populations. In Section \ref{sec:orbit}, we model the orbit of the stream and discuss its tidal disruption history. In Section \ref{sec:origin}, we discuss the nature of the 300S progenitor and infall scenarios. In Section \ref{sec:conclusion}, we summarize our results and present our conclusions.

\section{Stream Member Selection}
\label{sec:new_members}

\subsection{Stream Members in APOGEE-2}
\label{sec:apogee_members}

\par The Apache Point Observatory Galactic Evolution Experiment \citep[APOGEE;][]{majewski2015apache} survey obtained high signal-to-noise ratio ($\mbox{S/N}~\gtrsim 100$), high-resolution ($R \approx 22,500$) near-infrared ($1.51-1.70~\mu$m) spectra of 146,000 stars in and around the Milky Way with the Sloan Digital Sky Survey 2.5~m telescope \citep{gunn2006}. After processing with the APOGEE data reduction pipeline \citep{nidever2015}, stellar atmospheric parameters such as temperature, metallicity, and surface gravity are determined by the APOGEE Stellar Parameter and Chemical Abundances Pipeline \citep[ASPCAP;][]{garciaperez2016aspcap}. The APOGEE-2 project, part of the fourth-generation Sloan Digital Sky Survey \citep[SDSS-IV;][]{blanton2017sloan}, is extending APOGEE observations to a larger sample of stars (270,000 in DR14; \citealt{dr14}). This extended sample includes 9 dwarf galaxies out to distances of $\sim100$~kpc, and covers the southern hemisphere \citep{majewski2016,zasowski2017}. Several APOGEE-2 plug-plates span the location of the photometric features possibly associated with 300S, and provide the potential to study the chemical evolution history of the stream progenitor. 

\par
\begin{figure*}[t!]
\epsscale{1.2}
\plotone{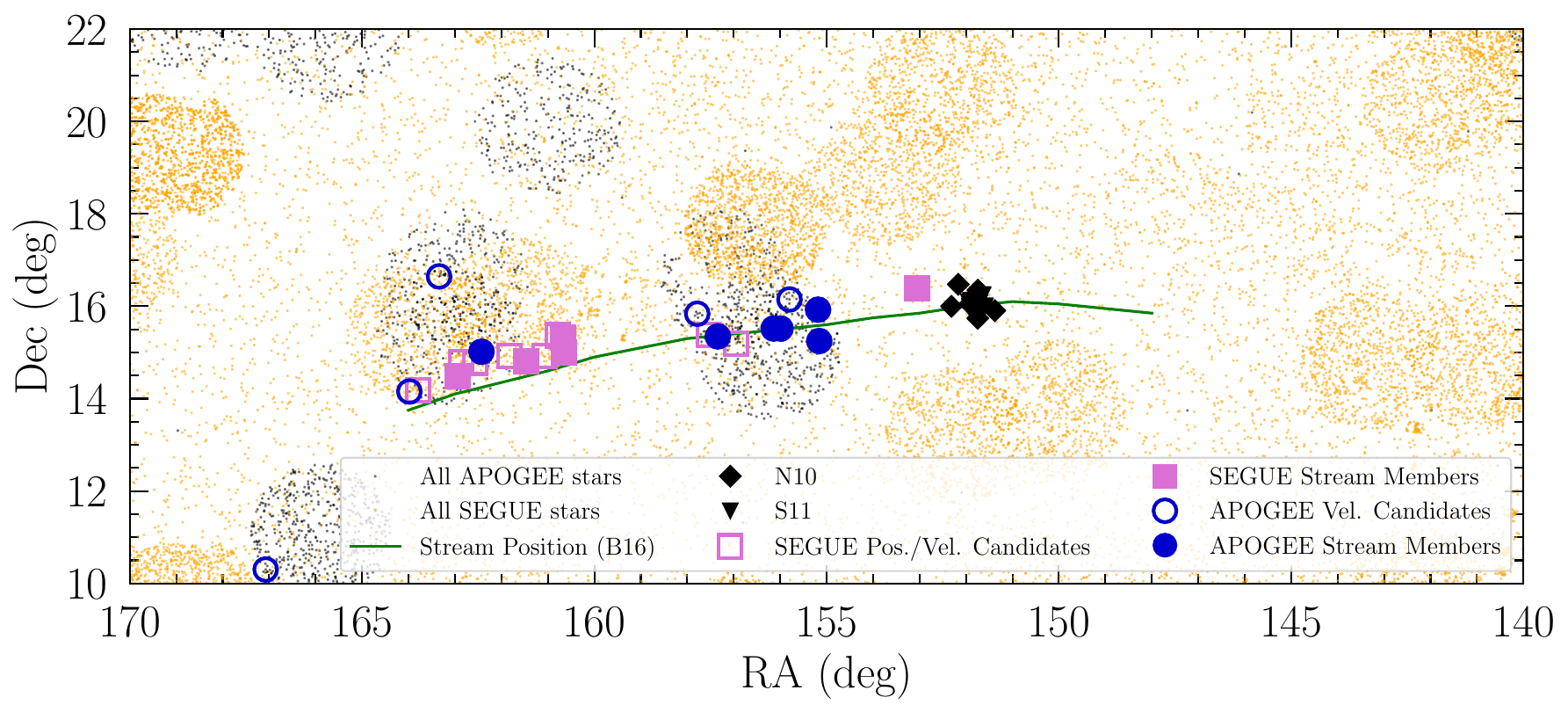}
\caption{Spatial distribution of previously known 300S members, as well as the 300S candidates identified in this study. We also include all of the stars from the APOGEE (black) and SEGUE (orange) surveys in this patch of sky for reference. B16 reported that the stream spatial extent is slightly larger than the plotted stream track. However, we only include this section of the track because that is where the presence of the stream is most apparent.}
\label{fig:mempos}
\end{figure*}

\begin{figure} 
\epsscale{1.2}
\plotone{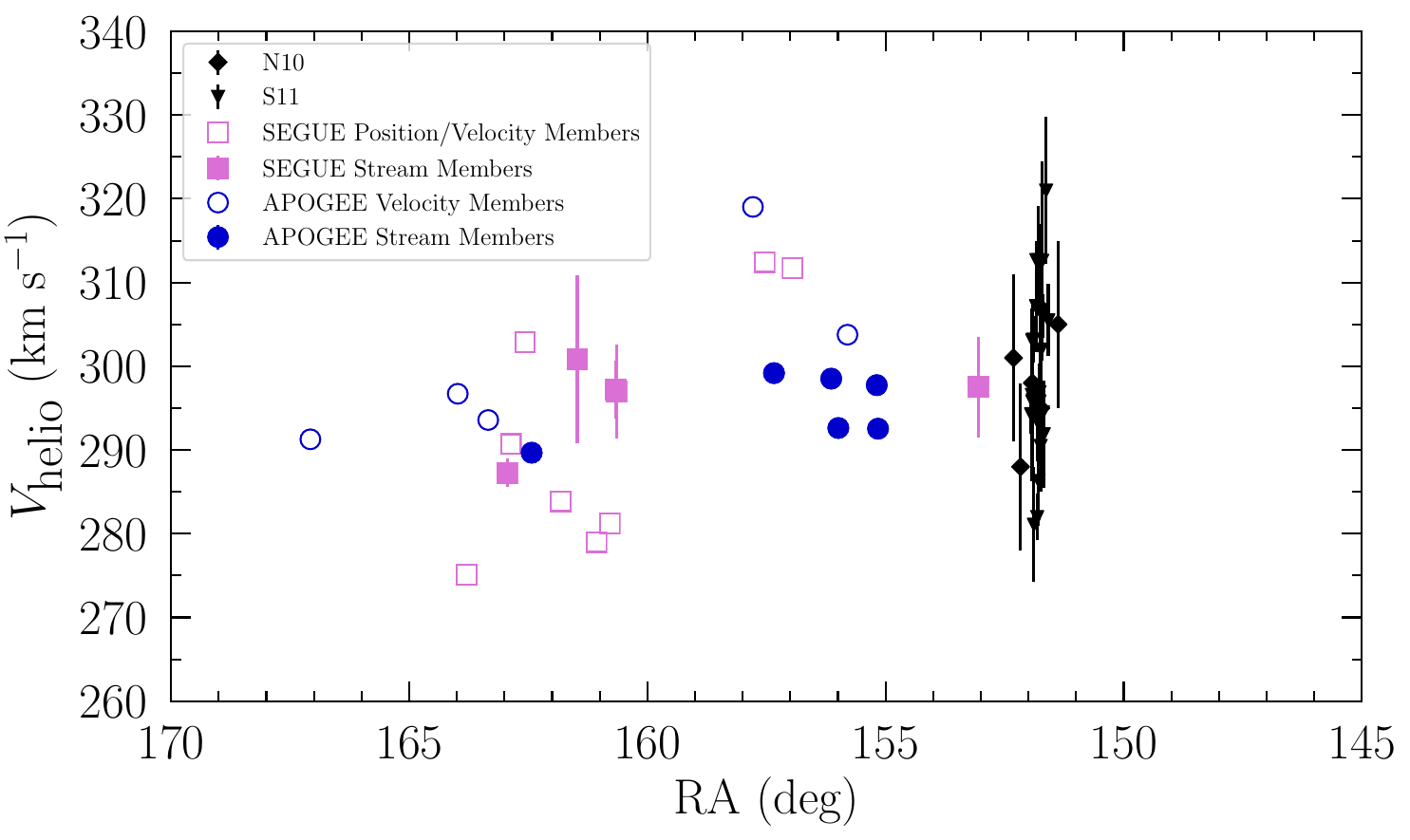}
\caption{Velocity distribution of stream members and candidates as a function of RA.  Since the stream runs approximately east-west, RA serves as a proxy for position along the stream. There may be a slight velocity gradient, with velocity decreasing as RA increases, but it is subtle.}
\label{fig:vel_ra}
\end{figure}

\begin{figure*}[ht!] 
\epsscale{1.2}
\plotone{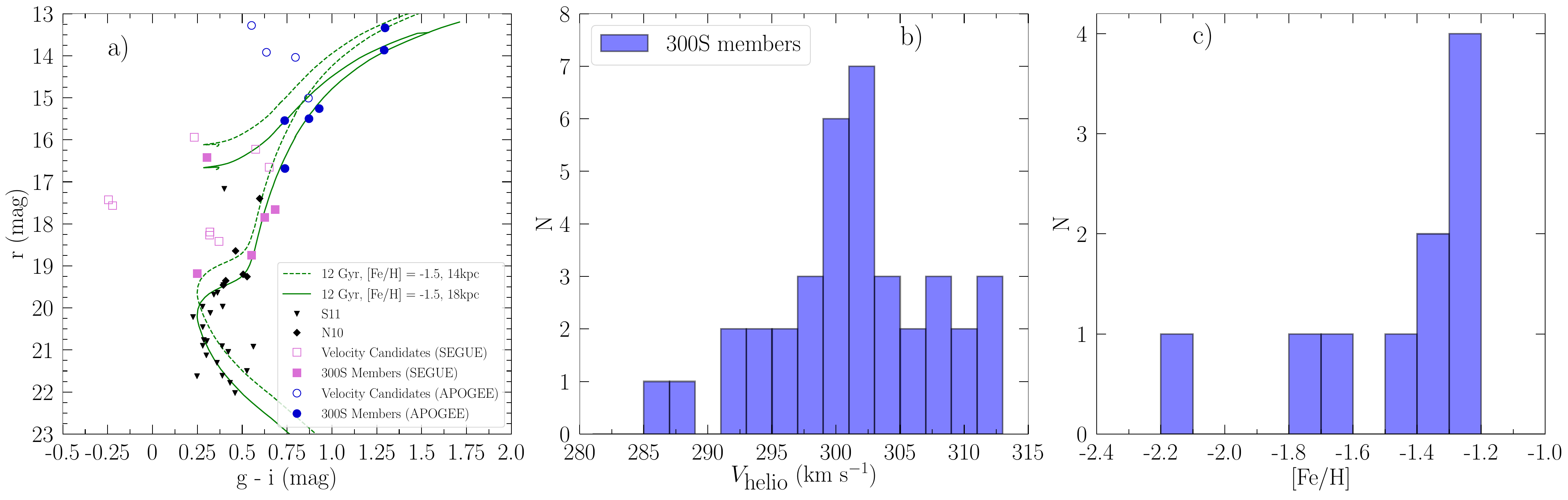}
\caption{(a) CMD of the 300S members and APOGEE-2 and SEGUE candidates. Most of the new stars lie on the red giant branch, with one horizontal branch member from SEGUE. The symbols are the same as in Figure~\ref{fig:mempos}; the hollow symbols are stars that passed the velocity criterion, but failed the CMD selection. (b) Velocity distribution of the stream members selected from the two surveys, as well as from S11 and N10. Bins are 2~\kms\ wide. (c) Metallicity distribution of stream members selected from the two surveys. Bins are 0.1 dex wide. There is a peak in metallicity at around $\mbox{[Fe/H]} = -1.4$, which is consistent with the metallicity measurement of one stream star by F13 and the photometric estimate from S11.}
\label{fig:memdistr}
\end{figure*}

 To search for potential 300S members, we begin with the APOGEE-2 allstars file, which was released in SDSS DR14 \citep{dr14}. This file includes the spectra of the original APOGEE survey, re-reduced and re-analyzed with the same software used on APOGEE-2 data. To isolate potential stream stars, we apply a velocity cut of $275~\mbox{\kms} < V_{\rm helio} < 325$~\kms, and a spatial cut of $140\degree < \alpha_{2000} < 170\degree$, $10\degree < \delta_{2000} < 22\degree$. We search in this region because it contains the photometric overdensity shown in Figure 3 from B16. Given the stream velocity dispersion measured by S11, the velocity criterion is quite generous ($3.5\sigma$), but it allows for a possible velocity gradient along the stream. As it turns out, none of the member stars we identify are near the edge of the velocity selection window, so the exact limits chosen do not affect our results. The stars resulting from this cut are shown in Figures \ref{fig:mempos}, \ref{fig:vel_ra} and \ref{fig:memdistr} as open blue circles.

\par In order to obtain a reference isochrone against which to compare potential 300S members, we fit the known stream population from S11 and N10. For these stars, as for the rest of the stars in this study, we obtain their Pan-STARRS DR1 \citep[PS1;][]{chambers2016pan} photometry\footnote{We use PS1 photometry because some APOGEE-2 candidates are bright enough to be saturated in SDSS images.}, using the mean PS1 magnitudes, and apply extinction corrections using the Bayestar17 3D dust map and extinction laws of \citet{green2018galactic}. A PARSEC isochrone \citep{marigo2017new} with $\mbox{[Fe/H]} = -1.5$ and an age of 12~Gyr at a distance of 18~kpc is a good match to the known stream population, in agreement with the results from S11 and F13.  

\par For the APOGEE-2 stars, we do an initial selection by examining their position on the color-magnitude diagram (CMD) shown in Figure \ref{fig:memdistr}a. B16 reported that the stream distance ranges from $14-19$~kpc. Because the stream distance near Segue~1 is 18~kpc, we posit that the stream distance increases toward smaller right ascension, and make a qualitative color selection based on identifying stars whose position on the CMD is close to, or in between, the best-fit isochrone shifted to distances of 14~kpc and 18~kpc. 

\par The star 2M10231170+1608483 lacks PS1 and SDSS photometry, so we could not confirm its membership using its position on the CMD. The stars 2M10310815+1550149, 2M10555533+1409147, and 2M10532191+1638441 lie well away from the stream isochrone; 2M10532191+1638441 also lies too far north of the stream trace to be considered a member. After this photometric and kinematic selection, the majority of the stars lie along the trace of the stream from B16 (Figure \ref{fig:mempos}), confirming for the first time that 300S is the kinematic counterpart of the photometric feature seen in B16. This association gives us an additional criterion for membership selection. In particular, the star 2M11081786+1018043 has colors consistent with it being a member of 300S, but falls several degrees beyond the trace of the stream from B16. It also falls outside of the track of the modeled orbit (see Section~\ref{sec:orbit}). For that reason, we also exclude it from 300S membership. 

\par In summary, out of the 11 APOGEE stars that remain after the velocity selection, we reject 5 of them as 300S members for the following reasons: 1 lacks photometry, 3 lie far from the isochrone shifted to the appropriate distance, and 1 falls beyond the trace of the stream as shown in B16. All of these stars are represented in Fig \ref{fig:mempos} and on the CMD in Fig \ref{fig:memdistr}a as blue circles.


\par We then proceed to a more quantitative selection for photometric stream members. We approximate the distance gradient along the stream as varying linearly with right ascension ($\alpha$), where $(d/1~\mbox{kpc}) = 48.9952 - 0.2083\alpha$ and $\alpha$ is in degrees. The $y$-intercept and slope were determined by using the RA and distance measurements from B16, where we assume $d_{\rm 300S} = 14$~kpc at $\alpha = 168\degr$, and $d_{\rm 300S} = 19$~kpc at $\alpha = 144\degr$. Within the APOGEE-2 stars, we select photometric members of the stream by requiring that members be within 0.12~mag of the theoretical isochrone at its corresponding distance along the stream. Although the photometric uncertainties for these bright stars are quite small, we allow for a relatively wide selection window around the isochrone in order to account for uncertainties in our distance model and the stellar populations of the stream. 

Table \ref{tab:apogeemem} provides the list of stars in APOGEE-2 that passed the spatial and velocity selection criteria. The stars that we consider members of 300S are marked with a ``1" under the ``MEM" field. Out of the 27 stars from APOGEE-2 in this patch of sky with $V_{\rm helio} > 250$~\kms, 6 are ultimately members of 300S. 

\subsection{Stream Members in SEGUE-1, SEGUE-2}
\label{sec:seguemembers}
\par The SEGUE (Sloan Extension for Galactic Understanding and Exploration) 1 \& 2 surveys collected $R \sim 1800$ spectra in the $3900-9000$~\AA\ wavelength range for $\sim 240,000$ stars with $14 < g < 20.3$ across a wide range of spectral types \citep{yanny2009segue}. The SEGUE data releases provide the radial velocity for every star, and, if the spectrum is of sufficient S/N, stellar atmospheric parameters such as metallicity, surface gravity, and effective temperature from the SEGUE Stellar Parameter Pipeline (SSPP) (\citealt{lee2008segue}, \citealt{lee2008segue2}, \citealt{allendeprieto2008segue}, \citealt{smolinski2011segue}, \citealt{lee2011alpha}).

\par To search for 300S members from the SEGUE surveys, we begin by selecting stars in the region displayed in Figure~3 from B16 with velocities between 275~\kms\ and 325~\kms, and with uncertainties of less than 10~\kms. To ensure that we do not have duplicate entries, we select stars that have the `scienceprimary' flag set to 1. The spatial distribution of SEGUE stars with velocities near 300~\kms\ does not obviously reveal the presence of the stream, so we make another spatial cut by selecting stars that are less than 0.75\degr\ away from the center of the stream trace shown in Figure~\ref{fig:mempos}. A total of 13 stars pass the velocity and position cuts, shown in Figures~\ref{fig:mempos}, \ref{fig:vel_ra} and \ref{fig:memdistr} as pink squares. 

\par In selecting for photometric members, we apply the same criterion described in Section \ref{sec:apogee_members}, leaving 6 candidates. One of the stars, PSO J104717.120+145503.948, should lie on the RGB of the stream at an extinction-corrected $r$-band magnitude of 16.65. However, its atmospheric parameters determined from SEGUE spectroscopy, $T_{\rm eff} = 5353$~K and $\log {g} = 4.4$, suggest that it is a main sequence star. Thus, we exclude this star from the member sample. 

\par The results of 300S member selection are shown in Figures~\ref{fig:mempos}, \ref{fig:vel_ra}, and \ref{fig:memdistr}. Table~\ref{tab:segmem} provides the list of stars in the SEGUE survey data that passed the spatial and velocity selection criteria. Table~\ref{tab:segabund} provides the atmospheric parameters of the same stars, obtained from the publicly available SEGUE data. It also contains [Fe/H], [$\alpha$/Fe], and [C/Fe] abundance ratios derived from the updated SSPP described in \citet[][$\alpha$-elements]{lee2011alpha} and \citet[][carbon]{lee2013carbon}. For these stars, the [Fe/H] measurements from the publicly available SEGUE data and from the updated SSPP pipeline are consistent with each other. In the rest of our analysis, we use the [Fe/H] measurements presented in Table 3. In both tables, the 5 stars that we consider members of 300S are marked with a ``1" under the ``MEM" field. 

\begin{deluxetable*}{lcccccccccccc}
\tablecaption{300S Candidate Members from APOGEE Data}
\tablenum{1}
\label{tab:apogeemem}
\tabletypesize{\footnotesize}
\tablehead{\colhead{APOGEE ID} & \colhead{RA} & \colhead{Dec} & \colhead{$V_{\rm helio}$} & \colhead{$\sigma_{V_{\rm helio}}$} & \colhead{[Fe/H]} & \colhead{$\sigma_{[Fe/H]}$\tablenotemark{a}} & \colhead{[C/Fe]} & \colhead{[C/Fe] (c.)\tablenotemark{b}} & \colhead{$g$} & \colhead{$r$} & \colhead{$i$} & \colhead{MEM} \\ 
\colhead{} & \colhead{($\degree$)} & \colhead{($\degree$)} & \colhead{(\kms)} & \colhead{(\kms)} & \colhead{} & \colhead{(dex)} & \colhead{} & \colhead{} & \colhead{(mag)} & \colhead{(mag)} & \colhead{(mag)} & \colhead{}}
\startdata
2M10203784+1514471 & 155.15769 & 15.24642 & 292.54 & 0.01 & $-$1.38   & 0.10 & $-$0.57 & $-$0.02 & 14.99  & 14.02 & 13.61 & 1 \\
2M10204415+1555327 & 155.18399 & 15.92577 & 297.74 & 0.06 & $-$1.24   & 0.10 & $-$0.69 & $-$0.45 & 16.02  & 15.36 & 15.03 & 1 \\
2M10235791+1530589 & 155.99132 & 15.51638 & 292.62 & 0.13 & ...       & ...  &     ... &     ... & 17.40  & 16.83 & 16.57 & 1 \\
2M10243358+1531009 & 156.13992 & 15.51694 & 298.52 & 0.04 & $-$1.28   & 0.10 & $-$0.50 & $-$0.33 & 16.25  & 15.60 & 15.31 & 1 \\
2M10292189+1520453 & 157.34122 & 15.34594 & 299.18 & 0.07 & $-$1.32   & 0.10 & $-$0.21 & $-$0.18 & 16.22  & 15.67 & 15.41 & 1 \\
2M10494291+1500530 & 162.42883 & 15.01473 & 289.68 & 0.01 & $-$1.28   & 0.10 & $-$0.65 & $-$0.10 & 14.21  & 13.39 & 12.89 & 1 \\
2M10231170+1608483 & 155.79876 & 16.14675 & 303.77 & 0.02 & $-$1.20   & 0.10 & $-$0.38 &     ... & ...    & 14.34 & 13.77 & 0 \\
2M10310815+1550149 & 157.78400 & 15.83748 & 319.03 & 0.05 & $-$1.00   & 0.10 & $-$0.08 &     ... & 14.71  & 14.11 & 13.87 & 0 \\
2M10532191+1638441 & 163.34129 & 16.64560 & 293.58 & 0.06 & $-$1.14   & 0.10 & $-$0.23 &     ... & 13.64  & 13.33 & 13.06 & 0 \\ 
2M10555533+1409147 & 163.98056 & 14.15408 & 296.72 & 0.09 & $-$1.39   & 0.10 & $-$0.72 &     ... & 14.43  & 13.97 & 13.77 & 0 \\
2M11081786+1018043 & 167.07442 & 10.30121 & 291.28 & 0.07 & $-$1.07   & 0.10 & $-$0.25 &     ... & 15.67  & 15.06 & 14.77 & 0 \\
\enddata
\tablenotetext{a}{Because ASPCAP uncertainties in [Fe/H] are unreasonably small (on the order of 0.01~dex), we assume a minimum $\sigma_{\rm [Fe/H]}$ of 0.1~dex for all the stars presented. For reference, \citet{holtzman2015abundances} note that the external uncertainties in metallicity from ASPCAP can range from  0.1~dex to 0.2~dex.}
\tablenotetext{b}{[C/Fe] abundances corrected for the depletion of [C/Fe] along the red giant branch using the methods in \citet{placco2014}.}
\tablecomments{The 11 APOGEE-2 stars that meet the velocity criteria in the region of the stream trace found by B16. For easier direct comparison to the PS1 catalog, the PS1 magnitudes presented in this table have not been corrected for extinction.}
\end{deluxetable*} 

\begin{deluxetable*}{lcccccccccc}
\tablecaption{300S Candidate Members from SEGUE Data}
\tablenum{2}
\label{tab:segmem}
\tabletypesize{\footnotesize}
\tablehead{\colhead{PS1 ID} & \colhead{RA} & \colhead{Dec} & \colhead{$V_{\rm helio}$} & \colhead{$\sigma_{V_{\rm helio}}$} & \colhead{[Fe/H]} & \colhead{$\sigma_{[Fe/H]}$} & \colhead{$g$} & \colhead{$r$} & \colhead{$i$} & \colhead{MEM}\\ 
\colhead{} & \colhead{($\degree$)} & \colhead{($\degree$)} & \colhead{(\kms)} & \colhead{(\kms)} & \colhead{} & \colhead{(dex)} & \colhead{(mag)} & \colhead{(mag)} & \colhead{(mag)} & \colhead{}} 
\startdata
PSO J101213.614+162336.187 & 153.05674 & 16.39340 & 297.5 & 6.0  & $-$2.07   & 0.08    & 18.25  & 17.76 & 17.51 & 1 \\
PSO J104236.584+150006.746 & 160.65240 & 15.00186 & 297.0 & 5.6  & $-$1.45   & 0.08    & 19.24  & 18.82 & 18.64 & 1 \\
PSO J104241.253+151913.286 & 160.67186 & 15.32034 & 297.2 & 3.4  & $-$1.76   & 0.08    & 18.42  & 17.95 & 17.73 & 1 \\
PSO J104552.952+144850.129 & 161.47066 & 14.81393 & 300.8 & 10.0 & $-$1.43   & 0.04    & 19.45  & 19.23 & 19.17 & 1 \\
PSO J105146.576+142850.068 & 162.94404 & 14.48055 & 287.3 & 1.7  & $-$1.41   & 0.05    & 16.79  & 16.52 & 16.42 & 1 \\
PSO J102747.821+151112.482 & 156.94925 & 15.18678 & 311.7 & 5.5  & $-$1.44   & 0.24    & 17.48  & 17.55 & 17.64 & 0 \\
PSO J103009.937+152241.212 & 157.54140 & 15.37809 & 312.4 & 5.3  & $-$1.57   & 0.14    & 17.64  & 17.69 & 17.78 & 0 \\
PSO J104309.200+152210.855 & 160.78830 & 15.36964 & 281.2 & 4.7  & $-$1.33   & 0.04    & 18.83  & 18.52 & 18.40 & 0 \\
PSO J104416.759+145459.165 & 161.06981 & 14.91639 & 279.0 & 6.1  & $-$1.23   & 0.02    & 18.60  & 18.34 & 18.24 & 0 \\
PSO J104717.120+145503.948 & 161.82127 & 14.91769 & 283.9 & 1.5  & $-$1.43   & 0.06    & 17.21  & 16.73 & 16.52 & 0 \\
PSO J105016.344+144644.466 & 162.56806 & 14.77903 & 302.9 & 2.8  & $-$1.26   & 0.01    & 16.96  & 16.30 & 16.34 & 0 \\
PSO J105127.367+144705.996 & 162.86398 & 14.78494 & 290.7 & 2.1  & $-$1.83   & 0.02    & 16.23  & 16.01 & 15.95 & 0 \\
PSO J105510.494+141100.053 & 163.79369 & 14.18334 & 275.1 & 6.4  & $-$2.48   & 0.01    & 18.54  & 18.27 & 18.17 & 0 \\
\enddata
\tablecomments{The stars from the SEGUE surveys that pass the spatial and velocity cut. For easier direct comparison to the PS1 catalog, the PS1 magnitudes presented in this table have not been corrected for extinction. The quoted metallicity uncertainties are SSPP internal errors; external uncertainties are 0.2~dex. PSO~J101213.614+162336.187 is rather metal-poor compared to the other members. However, its velocity is consistent with membership in 300S.}
\end{deluxetable*}

\subsection{Properties of Newly Identified Stream Members}

\par As illustrated in Figure \ref{fig:memdistr}a, most of the 300S members appear to lie on the red giant branch. However, there is one star from the SEGUE survey that lies on the horizontal branch, which may be advantageous for determining a more robust distance measurement of the stream at a position away from Segue~1. The mean metallicity of the stream from these measurements is $\mbox{[Fe/H]} = -1.48$. For the SEGUE stars, we adopt an external metallicity uncertainty of 0.2~dex. For the APOGEE-2 stars, we adopt an external metallicity uncertainty of 0.1~dex. Using those values, we calculate an intrinsic metallicity dispersion of 0.21$^{+0.12}_{-0.09}$~dex. \citet{holtzman2015abundances} note that the external uncertainty on ASPCAP [Fe/H] measurements ranges from 0.1~dex to 0.2~dex. The exact value that we adopt for the APOGEE-2 stars does not significantly affect our results because the metallicity dispersion is driven largely by the more metal-poor SEGUE stars. However, the choice of metallicity uncertainty for the SEGUE members does matter; a larger value can substantially reduce the derived intrinsic metallicity dispersion for 300S.  

\par Figure \ref{fig:vel_ra} shows the heliocentric velocity of the stream members as a function of RA. There may be a slight velocity gradient along the stream, with velocity increasing with negative RA, but it is subtle. A larger sample size would be needed to verify the existence of the gradient. 

\par For the 300S members in both the APOGEE-2 and SEGUE samples, we correct for the depletion of [C/Fe] along the red giant branch by applying the methods used in \citet{placco2014}. For the APOGEE-2 members, the respective carbon corrections for stars with available abundances (shown in the order in Table \ref{tab:apogeemem}) are 0.55, 0.24, 0.17, 0.03, and 0.55~dex. We present the corrected values in Table \ref{tab:apogeemem}. For the SEGUE stars, the corrected carbon abundances are included in Table~\ref{tab:segabund}. For these stars, the corrections are less than 0.1~dex. One 300S star, PSO~J104236.584+150006.746, can be classified as carbon-enhanced ($\mbox{[C/Fe]} = +0.90$). Given its metallicity, PSO~J104236.584+150006.746 is likely a CEMP-$s$ star enriched by a binary companion. Follow-up spectroscopy to measure its neutron-capture element abundances and velocity variability may be interesting to confirm this possibility. 

\begin{deluxetable*}{ccccccccccccc}
\tablecaption{Atmospheric Parameters and Abundances for 300S Candidates Members from SEGUE}
\tablenum{3}
\label{tab:segabund}
\tabletypesize{\footnotesize}
\tablehead{
\colhead{PS1ID} & \colhead{$T_{\mbox{eff}}$} & \colhead{$\sigma_{T_{\mbox{eff}}}$} & \colhead{$\log {g}$} & \colhead{$\sigma_{\log {g}}$} & \colhead{[Fe/H]} & \colhead{$\sigma_{\mbox{[Fe/H]}}$} & \colhead{[$\alpha$/Fe]} & \colhead{$\sigma_{\mbox{[$\alpha$/Fe]}}$} & \colhead{[C/Fe]} & \colhead{$\sigma_{\mbox{[C/Fe]}}$} & \colhead{[C/Fe] (c.)} & \colhead{MEM} \\
\colhead{} & \colhead{(K)} & \colhead{(K)} & \colhead{} & \colhead{(dex)} & \colhead{} & \colhead{(dex)} &  \colhead{} & \colhead{(dex)} & \colhead{} & \colhead{(dex)} & \colhead{} & \colhead{}}
\startdata
PSO J101213.614+162336.187 & 5057 &  83 & 2.3 & 0.1 & $-$2.17 & 0.13 & $+$0.85 & 0.17 & $-$0.26 & 0.12 & $-$0.26 & 1 \\
PSO J104236.584+150006.746 & 5367 &  47 & 2.6 & 0.1 & $-$1.65 & 0.11 & $+$0.28 & 0.22 & $+$0.88 & 0.11 & $+$0.90 & 1 \\
PSO J104241.253+151913.286 & 5216 &  63 & 2.5 & 0.1 & $-$1.75 & 0.13 & $+$0.39 & 0.15 & $+$0.04 & 0.10 & $+$0.06 & 1 \\
PSO J104552.952+144850.129 & 6557 &  53 & 4.0 & 0.6 & $-$1.26 & 0.20 & $+$0.21 & 0.20 & $+$0.57 & 0.42 & $+$0.57 & 1 \\
PSO J105146.576+142850.068 & 6131 &  62 & 2.3 & 0.3 & $-$1.45 & 0.07 & $+$0.43 & 0.08 & $-$0.12 & 0.22 & $-$0.10 & 1 \\
PSO J102747.821+151112.482 & 8275 &  22 & 4.1 & 0.2 & $-$1.56 & 0.11 &     ... &  ... &     ... &  ... &     ... & 0 \\
PSO J103009.937+152241.212 & 7970 &  76 & 4.0 & 0.1 & $-$1.49 & 0.08 &     ... &  ... & $+$2.19 &  0.5 & $+$2.20 & 0 \\
PSO J104309.200+152210.855 & 6047 &  55 & 4.2 & 0.1 & $-$1.30 & 0.02 & $+$0.31 & 0.16 & $+$0.21 & 0.14 & $+$0.21 & 0 \\
PSO J104416.759+145459.165 & 6313 &  53 & 3.3 & 0.2 & $-$1.27 & 0.11 & $+$0.26 & 0.20 & $+$0.58 & 0.28 & $+$0.59 & 0 \\
PSO J104717.120+145503.948 & 5353 &  17 & 4.4 & 0.0 & $-$1.44 & 0.04 & $+$0.46 & 0.04 & $+$0.04 & 0.02 & $+$0.04 & 0 \\
PSO J105016.344+144644.466 & 6590 & 105 & 3.2 & 0.2 & $-$1.46 & 0.07 & $+$0.67 & 0.10 & $+$0.46 & 0.16 & $+$0.48 & 0 \\
PSO J105127.367+144705.996 & 6453 &  32 & 3.7 & 0.1 & $-$1.87 & 0.06 & $+$0.54 & 0.14 & $<$0.01 &  ... &     ... & 0 \\
PSO J105510.494+141100.053 & 6329 &  56 & 3.8 & 0.2 & $-$2.75 & 0.12 & $+$0.64 & 0.15 & $+$1.66 & 0.42 & $+$1.66 & 0 \\
\enddata
\tablecomments{Atmospheric parameters and [Fe/H], [$\alpha$/Fe], and [C/Fe] abundances for the 11 SEGUE stars that pass the spatial and velocity cuts. $T_{\mbox{eff}}$ and $\log {g}$ were obtained from the publicly available SEGUE data, while the abundance measurements were obtained using the updated SSPP, described in \citet[][$\alpha$-elements]{lee2011alpha} and \citet[][carbon]{lee2013carbon}. The column ``[C/Fe] (c.)" provides the carbon abundances of these stars, corrected for evolutionary stage using the methods described in \citet{placco2014}. The uncertainties quoted are SSPP internal errors. External uncertainties in $T_{\rm eff}$, $\log {g}$, [Fe/H], [$\alpha$/Fe] and [C/Fe] are 125~K, 0.35~dex, 0.2~dex, 0.2~dex and 0.25~dex respectively. The star PSO~J104236.584+150006.746 can be classified as carbon-enhanced ($\mbox{[C/Fe]} = +0.90$); given its metallicity, it is likely a CEMP-$s$ star.}
\end{deluxetable*}

\section{Chemical Abundance Analysis}
\label{sec:abundances}

In this section we examine the chemical abundances of the stream members identified in Section~\ref{sec:new_members}. We analyze detailed abundances for the six APOGEE-2 stream members identified in Section~\ref{sec:apogee_members}. For one of these stars, 2M10235791+1530589, the ASPCAP pipeline was unable to determine any abundances.  However, a line-by-line analysis of its spectrum suggests similar abundances to the other five stars. We also examine the [$\alpha$/Fe] abundance ratios for the SEGUE stream members. 

\subsection{Comparison Sample Selection}

\par From the APOGEE-2 dataset, we select various other sets of stars with which to compare chemical abundances. For all stars, we verify that none have the `STAR\_BAD' ASPCAP flag, which encodes any unreliable ASPCAP measurements, set \citep{holtzman2015abundances}. We also ensure that these stars have internal uncertainties less of than 0.2 dex for each element considered. 

\par For a dwarf galaxy comparison sample, we select members of the Sagittarius (Sgr) dwarf galaxy from APOGEE (see \citealt{hasselquist2017apogee} for a detailed study of the chemical abundances of Sagittarius). We begin by selecting stars with the `APOGEE\_SGR\_DSPH' flag set in the APOGEE\_TARGET1 column. From that sample, we apply a velocity cut, selecting stars with $120~\mbox{\kms} < V_{\rm helio} < 160$~\kms, which isolates most stars in the core and removes some potential contaminants.  For data on other dwarf galaxies, we use the measurements of \citet{shetrone2003vlt} for Sculptor, Leo I, Carina and Fornax, \citet{cohen2009chemical} for Draco, and \citet{cohen2010chemical} for Ursa Minor.

\par We also select APOGEE stars in the globular clusters M13 and M92 for comparison. We chose M13 because its metallicity is similar to that of the stream, and M92 as a metal-poor reference. We take cluster membership information from \citet{meszaros2015exploring}. However, we use chemical abundances from ASPCAP to control for potential systematics due to different analysis methods. 

\par We construct our halo sample by obtaining distances from the APOGEE DR14-Based Distance Estimations value added catalog, which was constructed using the isochrone matching technique \citep[NICE;][]{schultheis2014extinction}. To determine the height above the Galactic plane where 90\% of the stars are from the halo as a function of Galactocentric radius, we use the Trilegal model \citep*{vanhollebeke2009stellar}. This height varies from ~3 to 10~kpc as a function of radius.

\par We also supplement our position-selected halo sample with stars that have 3D space motions consistent with halo membership, which we base on Gaia DR1 proper motions (see \citealt{gaia2016gaia}, \citealt{brown2016gaia}, \citealt{lindegren2016astrometry}, and \citealt{arenou2017gaia}, among others) and APOGEE DR14 radial velocities. We select our stars as those that have rotational and $UW$ velocities that differ from those of thick- and thin-disk stars by more than 2$\sigma$. Due to the magnitude limits in Gaia DR1, there was very little overlap between these two differently selected samples. 

\subsection{Light-Element Abundance Correlations}

\begin{figure*}
\epsscale{1.2}
\plotone{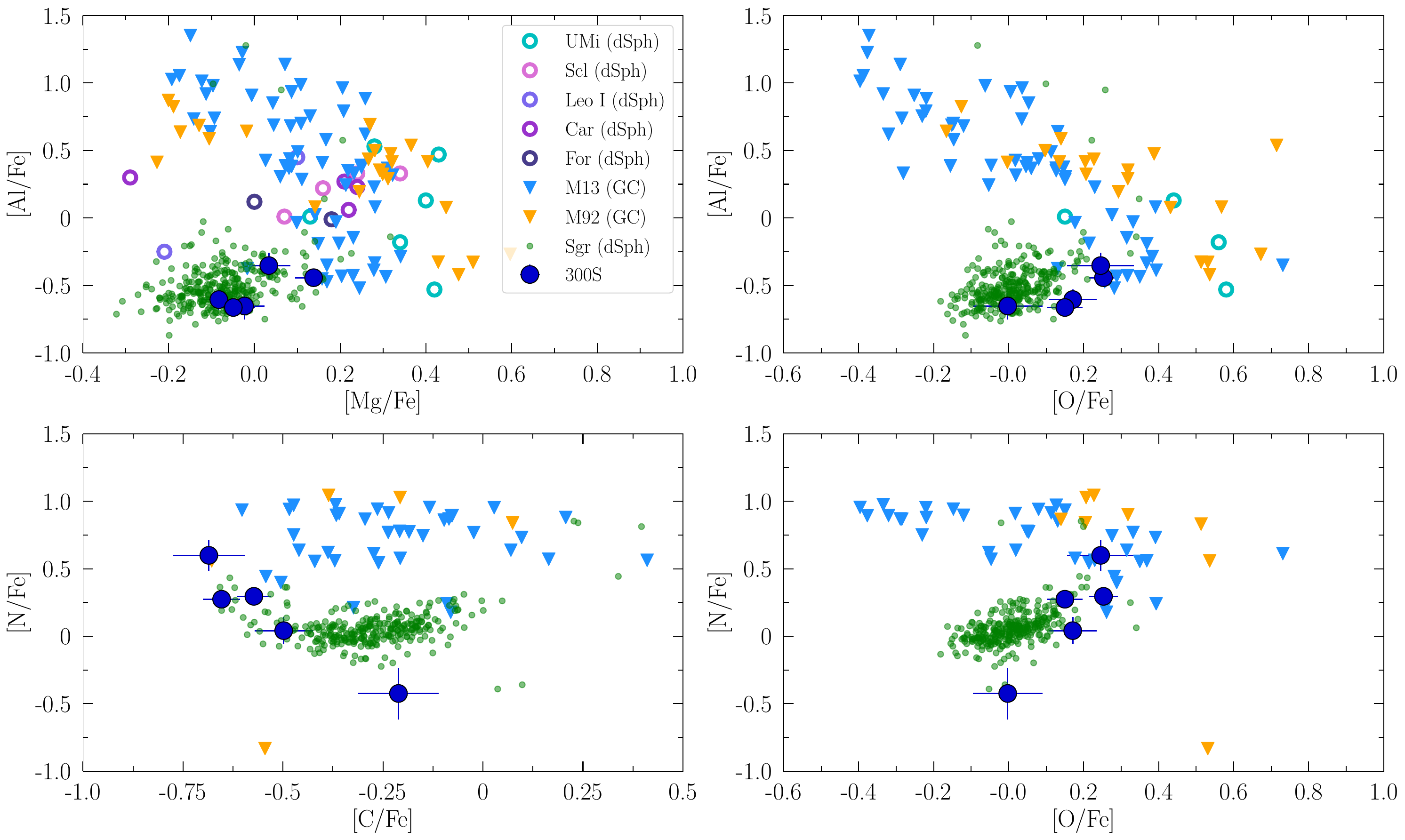}
\caption{Light-element abundance ratios of 300S stars (filled blue circles) in comparison to globular clusters and dwarf galaxies. Data for UMi are from \cite{cohen2010chemical}. Data for the other dwarf spheroidals (dSphs), not including Sgr, are from \cite{shetrone2003vlt}.  300S does not follow the light-element correlations seen in globular clusters. Instead, its abundance pattern resembles Sgr.}
\label{fig:chem_corr}
\end{figure*}

\par We begin our investigation into the nature of the stream progenitor by comparing its chemical abundance pattern to those of globular clusters. Globular cluster stars obey well-known correlations between abundances of light elements including carbon, nitrogen, oxygen, sodium, aluminum, and magnesium \citep[see, e.g.,][and references therein]{gratton2004}. In Figure \ref{fig:chem_corr} we examine the light-element abundances of 300S in comparison to M92 and M13, as well as to Local Group dwarf galaxies. Because Na is too weak to be measured in the $H$-band at low metallicity ($[\mathrm{Na/H}] \lesssim -1$), we could not test for a Na-O anti-correlation among the 300S stars.

\par It is apparent that members of 300S do not display the chemical abundance correlations of globular clusters, either in the direction of the correlation or the shape of the distribution. From comparison with Figure 9 of \citet{meszaros2015exploring}, which shows the Mg-Al correlations of the globular clusters from that study, we note in particular that 300S does not resemble either the first-generation or second-generation stars in M13. The chemical abundances of 300S are also much more similar to those of the Sagittarius dwarf galaxy than either of the globular clusters. This comparison strongly suggests that the progenitor of 300S is more likely to be a dwarf galaxy than a globular cluster. 

\subsection{[C/N] Ratio}

\par
\begin{figure*}
\epsscale{1.2}
\plotone{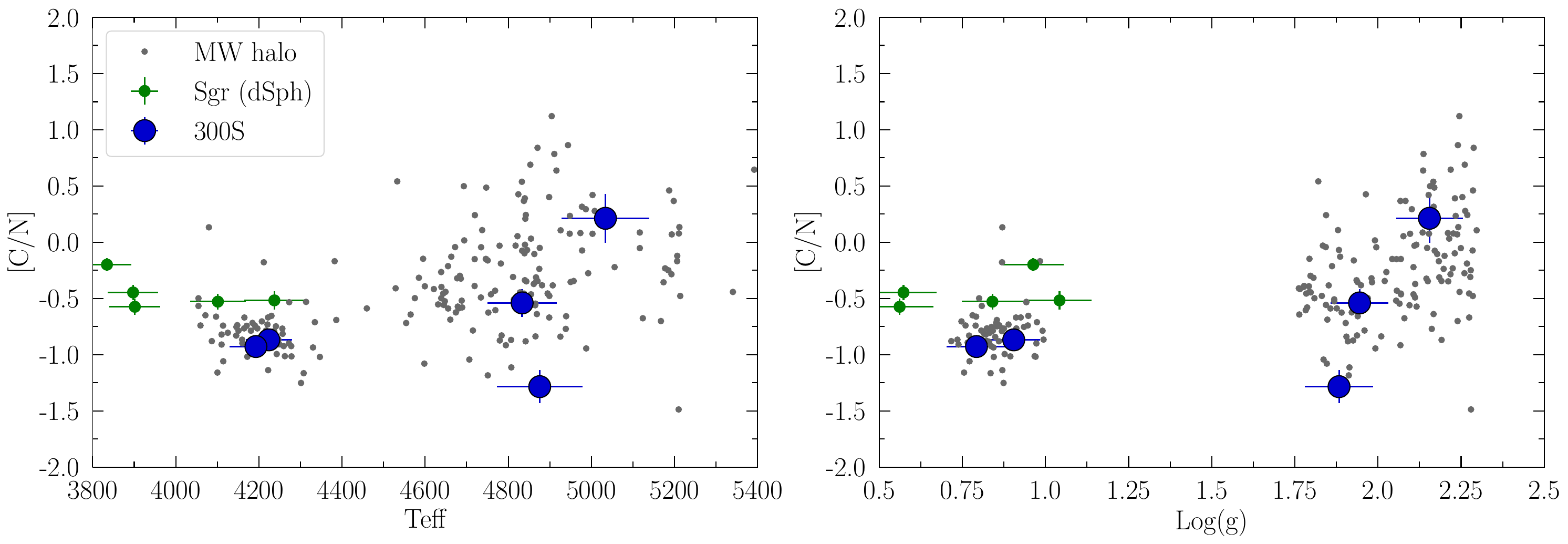}
\caption{Comparison of the [C/N] ratio of 300S stars to that of Sgr and MW halo stars with similar metallicity and $\log {g}$. In the range of $-1.5 < \mbox{[Fe/H]} < -1.0$, 300S looks similar to both Sgr and the MW halo, suggesting that they are approximately of the same age.}
\label{fig:cnlogg}
\end{figure*}

 To first order, the [C/N] ratio serves as an indicator of age. In Figure \ref{fig:cnlogg} we compare the [C/N] ratios of 300S to those of Sagittarius and the Milky Way halo. To control for evolutionary stage and possible deviations from local thermodynamic equilibrium (LTE), we select stars from our comparison sample that have similar $\log {g}$ and [Fe/H] to 300S.  In the regime of $-1.5 < \mbox{[Fe/H]} < -1.0$, the 300S members appear to be approximately as old as Sgr and the halo.

\subsection{Chemical Abundance Patterns}

\par To control for non-LTE effects, we originally only selected stars that are similar to the 300S sample in $T_{\rm eff}$ and $\log{g}$. However, the conclusions we drew from considering only such $T_{\rm eff}$ and $\log{g}$ ``twins'' were the same as those from considering the full halo sample. Therefore, we present our full sample of halo stars in the following figures in order to better illustrate the halo chemical abundance distribution, and highlight the stars that are $T_{\rm eff}$ and $\log{g}$ twins of 300S.

\par
\begin{figure}
\epsscale{1.2}
\plotone{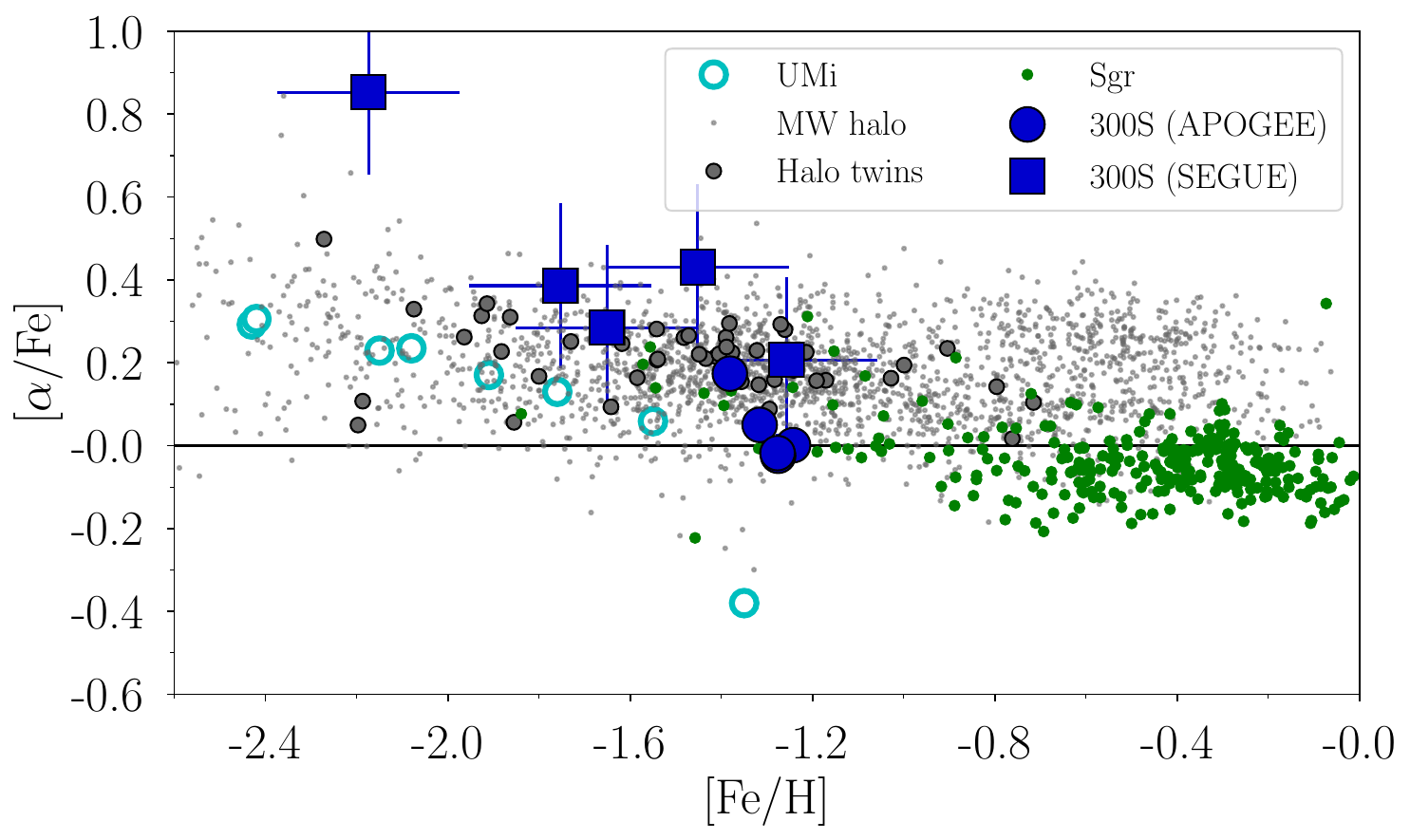}
\caption{Comparison of [$\alpha$/Fe] ratio of 300S stars to that of Sgr and MW halo stars, as well as to Ursa Minor. MW halo stars that are $T_{\rm eff}$ and $\log{g}$ twins of 300S are shown as solid gray circles with black outlines. The error bars on the SEGUE data points are based on their external uncertainties. The stars in 300S appear to reach solar levels of [$\alpha$/Fe] at around [Fe/H] = $-1.3$, which is between the respective metallicities where Ursa Minor and Sagittarius reach solar [$\alpha$/Fe] levels.}
\label{fig:alphageneral}
\end{figure}

 Figure \ref{fig:alphageneral} compares the [$\alpha$/Fe] abundance ratios of 300S to those of Sgr, the MW halo, and Ursa Minor. To ensure a consistent comparison to the [$\alpha$/Fe] ratios for the SEGUE stars, we calculate [$\alpha$/Fe] for the other populations by taking a weighted average abundance of Mg, Ca, Ti, and Si, where the weights for these four elements are given in \citet{lee2011alpha}. For Ti, we give the same weight to the abundances of its different ionization states. Ursa Minor was the only dSph from the literature included in this sample, because it 
alone has published abundances for all four of the above-mentioned elements. The ``knee" of the [$\alpha$/Fe] ratio as a function of metallicity \citep{tinsley1979stellar} for 300S appears to occur at around [Fe/H] = $-1.3$. 300S reaches solar [$\alpha$/Fe] at a similar metallicity to the classical dSphs.

\par
\begin{figure*}
\epsscale{1.2}
\plotone{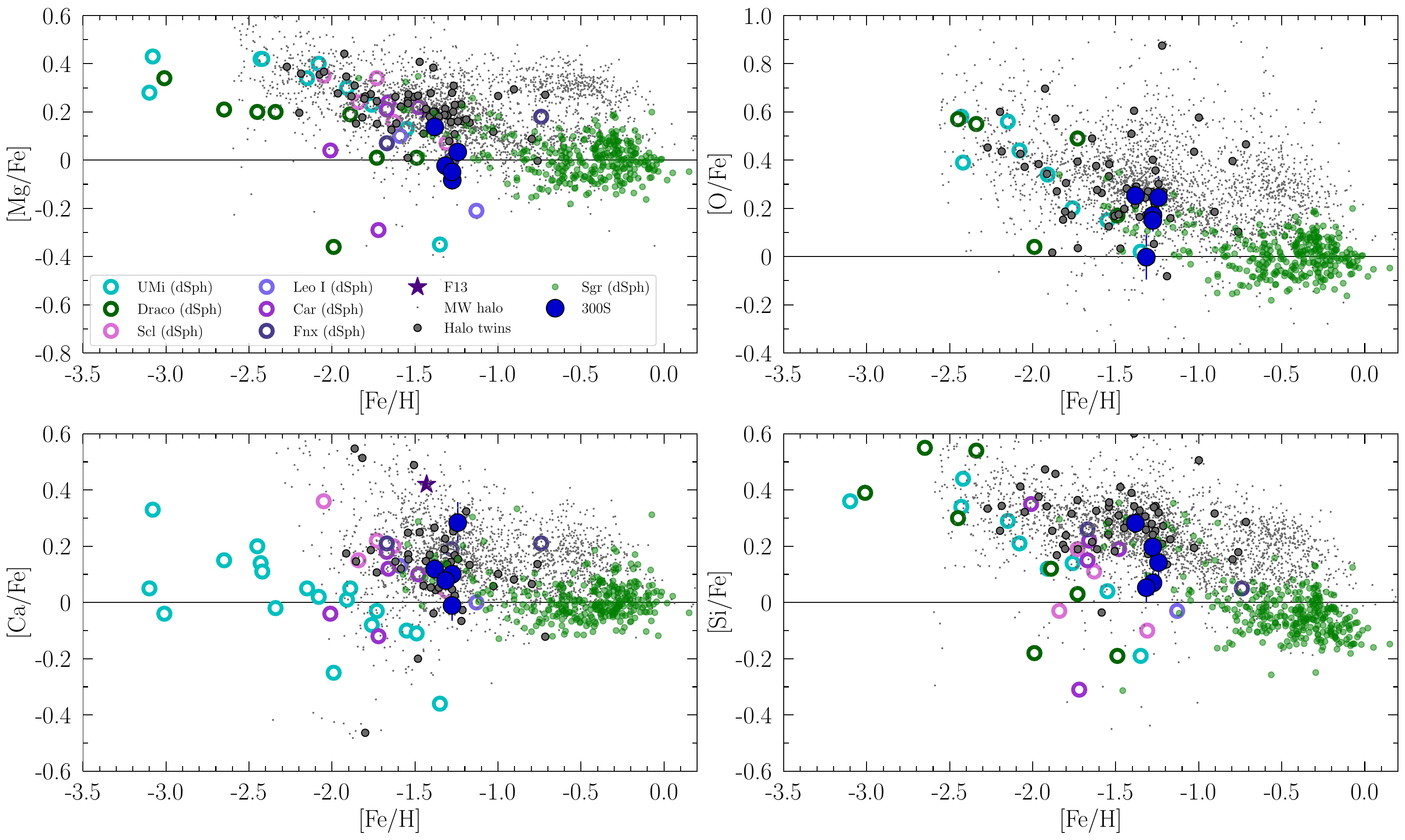}
\caption{The $\alpha$-element abundance ratios as a function of metallicity for 300S stars compared to those of the halo and other Milky Way satellites. Data for UMi and Draco are from \cite{cohen2010chemical} and \cite{cohen2009chemical}, respectively. Data for the other dSphs, not including Sgr, are from \cite{shetrone2003vlt}. MW halo stars that are $T_{\rm eff}$ and $\log{g}$ twins of 300S are shown as solid gray circles with black outlines. We also include the 300S member analyzed in F13, shown as the purple star when relevant. Overall, 300S displays similar $\alpha$-element abundances to those of the Local Group dwarf galaxies and the MW halo.}
\label{fig:alpha}
\end{figure*}

 Figure \ref{fig:alpha} compares individual $\alpha$-elements as a function of metallicity in 300S and Sgr. The $\alpha$-element abundances of 300S generally match those of the classical dSphs. 300S may be slightly enriched in calcium compared to Sgr and the other dwarf spheroidals. Compared to the MW halo, 300S also seems to have similar $\alpha$-element abundances. 

\par
\begin{figure*}
\epsscale{1.2}
\plotone{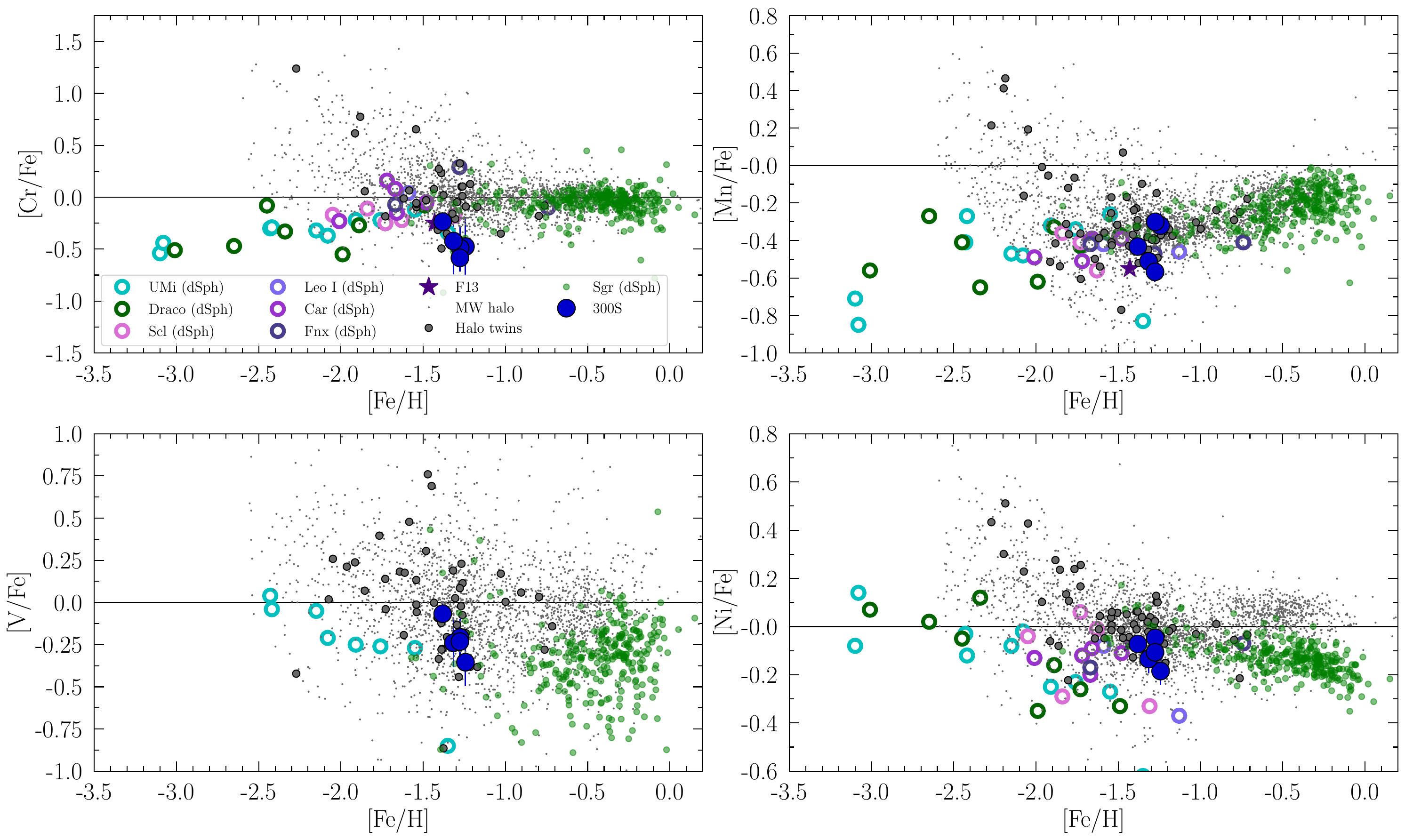}
\caption{The Fe-peak element abundance ratios for 300S stars relative to those of Milky Way systems. Data for other dwarf galaxies in these plots are from the same studies cited previously. MW halo stars that are $T_{\rm eff}$ and $\log{g}$ twins of 300S are shown as solid gray circles with black outlines.}
\label{fig:fe_peak}
\end{figure*}

 Figure \ref{fig:fe_peak} compares the abundances of Fe-peak elements in 300S to those of the other Milky Way systems. 300S appears to be deficient in Cr relative to the other dwarf spheroidals. Within the uncertainties, 300S has similar Mn and Ni abundances to the dwarf galaxies.  Overall, 300S has similar Fe-peak abundance patterns relative to the MW halo. Compared to the reference globular clusters, 300S is deficient in Fe-peak elements. 

\par That 300S has similar chemical abundance patterns to the Milky Way halo, as well as dSphs, and different ones from those of globular clusters, further suggests that the 300S progenitor is a dwarf galaxy. 

\section{Tidal Disruption History}
\label{sec:orbit}
\subsection{Proper Motion Modeling}

In Section~\ref{sec:new_members}, we determined the path of 300S along the sky and its distance and velocity as a function of position.  In order to calculate the orbit of the stream around the Milky Way, we also require proper motions.

We initially attempted to employ proper motion catalogs such as UCAC5 \citep{zacharias2017ucac5}, and the recently released GPS-1 \citep{tian2017gaia}, to constrain the proper motions of the stream members. However, we found that the published proper motions of the APOGEE-2 and SEGUE stars in the stream exhibit a large scatter, and the UCAC5 and GPS-1 proper motions are not in very good agreement for the stars with measurements in both catalogs. Table~\ref{tab:pm} presents the catalog proper motion values for the members of 300S identified in Section~\ref{sec:new_members}. 

\begin{deluxetable*}{cccccccccc}
\tablecaption{Proper motions for 300S Members from UCAC5, GPS-1}
\tablenum{5}
\tabletypesize{\footnotesize}
\tablehead{
\colhead{StarID} & \colhead{$\mu_{\alpha,GPS1}\cos{\delta}$} & \colhead{$\sigma$} & \colhead{$\mu_{\delta,GPS1}$} & \colhead{$\sigma$} & \colhead{$\mu_{\alpha,UCAC5}\cos{\delta}$} & \colhead{$\sigma$} & \colhead{$\mu_{\delta,UCAC5}$} & \colhead{$\sigma$} \\
\colhead{} & \colhead{(\masyr)} & \colhead{(\masyr)} & \colhead{(\masyr)} & \colhead{(\masyr)} & \colhead{(\masyr)} & \colhead{(\masyr)} & \colhead{(\masyr)} & \colhead{(\masyr)}}
\startdata
2M10203784+1514471               & -5.7 & 3.0 & -2.7 & 2.5 & ...  & ...  & ...  & ... \\
2M10204415+1555327               & -5.7 & 1.7 & -1.4 & 1.3 & ...  & ...  & ...  & ... \\
2M10235791+1530589               & -0.4 & 1.4 & 0.0  & 1.2 & ...  & ...  & ...  & ... \\
2M10243358+1531009               & -2.4 & 1.3 & -7.7 & 1.2 & -6.5 & 3.8  & -3.4 & 2.7 \\
2M10292189+1520453               & 0.0  & 1.5 &  0.2 & 1.2 & ...  & ...  & ...  & ... \\
2M10494291+1500530               & 0.0  & 3.0 & -8.3 & 2.6 & -4.8 & 1.1  & -4.5 & 1.1 \\
PSO J101213.614+162336.187 & -1.9 & 1.7 & -1.8 & 1.2 & ...  & ...  & ...  & ... \\
PSO J104236.584+150006.746 & -3.5 & 1.9 & -2.7 & 1.5 & ...  & ...  & ...  & ... \\ 
PSO J104241.253+151913.286 & -6.4 & 1.8 & -6.1 & 1.8 & ...  & ...  & ...  & ... \\
PSO J104552.952+144850.129 & -3.4 & 2.1 & -1.6 & 1.7 & ...  & ...  & ...  & ... \\
PSO J105146.576+142850.068 & -1.5 & 1.4 & -4.4 & 1.4 & -8.9 & 14.9 & 1.0  & 9.7 \\
\enddata
\label{tab:pm}
\tablecomments{Proper motions for the new 300S stars, obtained from the GPS-1 and UCAC5 catalogs. Directly to the right of every proper motion measurement is its corresponding measurement uncertainty.}
\end{deluxetable*}

\par Thus, we instead infer the proper motion of the stream by considering many possible orbits subject to the constraints of its known properties. We test a grid covering the full range of plausible proper motions to see which of them produce a stream with (1) the observed path, constrained by B16, (2) the distance along the stream, which we were able to obtain for APOGEE-2 stars from \cite{queiroz2018starhorse}, and (3) the velocity along the stream, constrained by the APOGEE-2 and SEGUE 300S members. 

\par We model 300S using the open-source code \textit{galpy} \citep{bovy2015galpy}, and approximate the orbit of the stream using point-particle integration. We integrate the orbits in the \texttt{MWPotential2014} potential, which is the standard Milky Way model in \textit{galpy}, and adopt the solar motion from \citet{schoenrich2010}. In order to initialize an orbit, \textit{galpy} requires 6D phase space information about the point where the orbit is initialized. We initialize the orbit of 300S at the location of the S11 stars (i.e., Segue~1) because that region has the best-constrained values as a result of the large sample of confirmed main sequence member stars. At that location, $\alpha = 151.8\degr$, $\delta = 16.1\degr$, $V_{\rm helio} = 298.8$~\kms, and $D_{\odot} = 18$~kpc (S11). 

\begin{figure}[!ht] 
\includegraphics[scale=0.5]{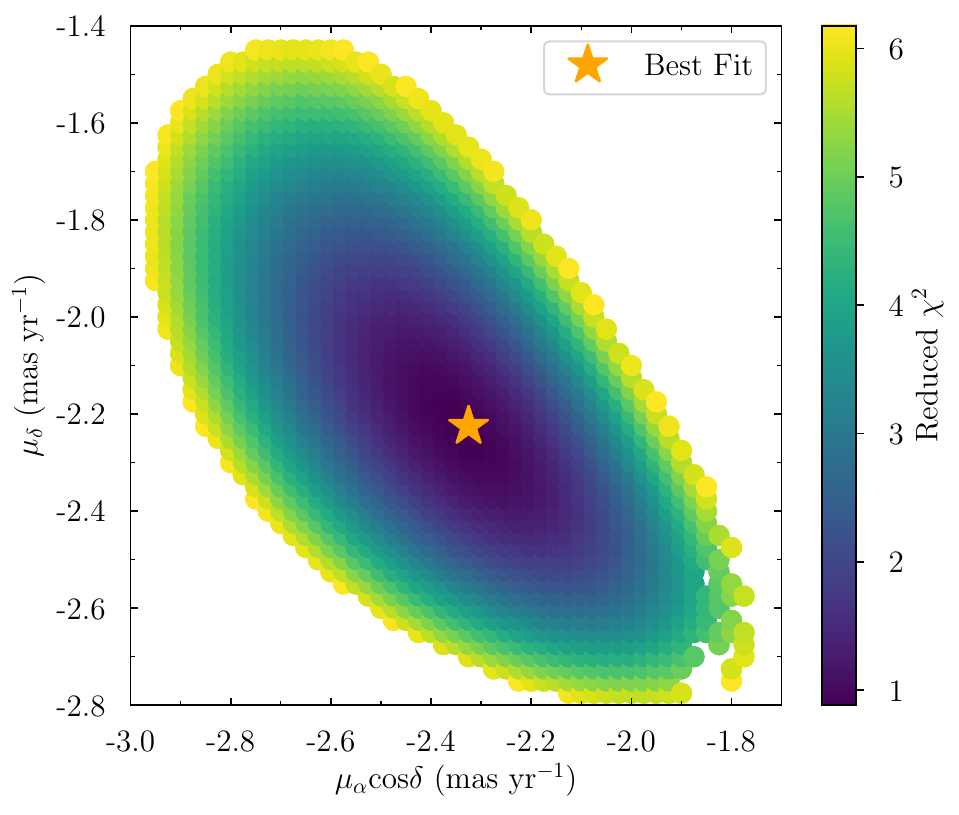}
\caption{Reduced $\chi^{2}$ value of the best-fitting orbit (DOF=72). The best-fit $\mu_{\alpha}\cos{\delta}$, $\mu_{\delta}$ values are $-$2.33~\masyr and $-$2.22~\masyr, respectively. The orbit from those proper motions has a reduced $\chi^{2}$ value close to 1, suggesting that the orbit model is a good fit to the data.}
\label{fig:pmfit}
\end{figure}

\par For every orbit corresponding to a proper motion, we compute its $\chi^{2}$ value based on its fit to the stream observables. Since the velocity errors for APOGEE are very small \citep{nidever2015}, we calculate $\chi^2$ for the APOGEE velocities assuming a Gaussian dispersion of 3 \kms 1 as a reasonable velocity dispersion for a stellar stream (\citealt{newberg2010orphan}, \citealt{casey2013orphan}). The velocity dispersion of the stream further away from Segue 1 is not well-constrained (see Section \ref{sec:origin}), so we allow for the possibility of a velocity dispersion without deviating too far from the measured values. We estimate the width of the stream to be 0.94~degree, and, assuming that our points are roughly centered on the stream, adopt a Gaussian dispersion of 0.47~degree for our on-sky spatial uncertainty.

\par Figure \ref{fig:pmfit} shows the results of the $\chi^{2}$ fits for a region around the best fit proper motion. The proper motion of the orbit corresponding to the lowest $\chi^{2}$ fit is $\mu_{\alpha}\cos{\delta} = -$2.33~\masyr, and $\mu_{\delta} = -$2.22~\masyr. The reduced $\chi^{2}$ value of the orbit fit at those proper motions, with a DOF of 72, is close to 1, suggesting that the model is a good fit. This proper motion is also close to the weighted mean of the GPS-1 proper motions of the member stars ($\mu_{\alpha}\cos{\delta} = -2.5 \pm 0.5$~\masyr, $\mu_{\delta} = -2.8 \pm 0.4$~\masyr), despite the large uncertainties of many of the individual measurements. 

\par For completeness, we fit an orbit with a positive $\mu_{\alpha}$, where the stream would travel in the opposite direction, and obtain proper motions of $\mu_{\alpha}\cos{\delta} = 1.48$~\masyr, and $\mu_{\delta} = -2.63$~\masyr. This orbit has a distance vs. $RA$ gradient that points in the opposite direction of the negative $\mu_{\alpha}$ solution. This orbit also has a corresponding reduced $\chi^2$ value of 2.15, suggesting that the positive $\mu_{\alpha}$ orbit is a poorer fit to the data, and is on a lower eccentricity orbit with a larger perigalactic distance of 17~kpc.

\par Thus, we prefer the negative $\mu_{\alpha}$ solution due to its better agreement with the GPS-1 proper motions, and because its much smaller perigalactic distance is more consistent with the observed disruption of the stream progenitor.  The forthcoming Gaia DR2 measurements will eliminate this degeneracy.

\subsection{Properties of the Modeled Orbit}

\begin{figure}[!ht] 
\includegraphics[scale=0.34]{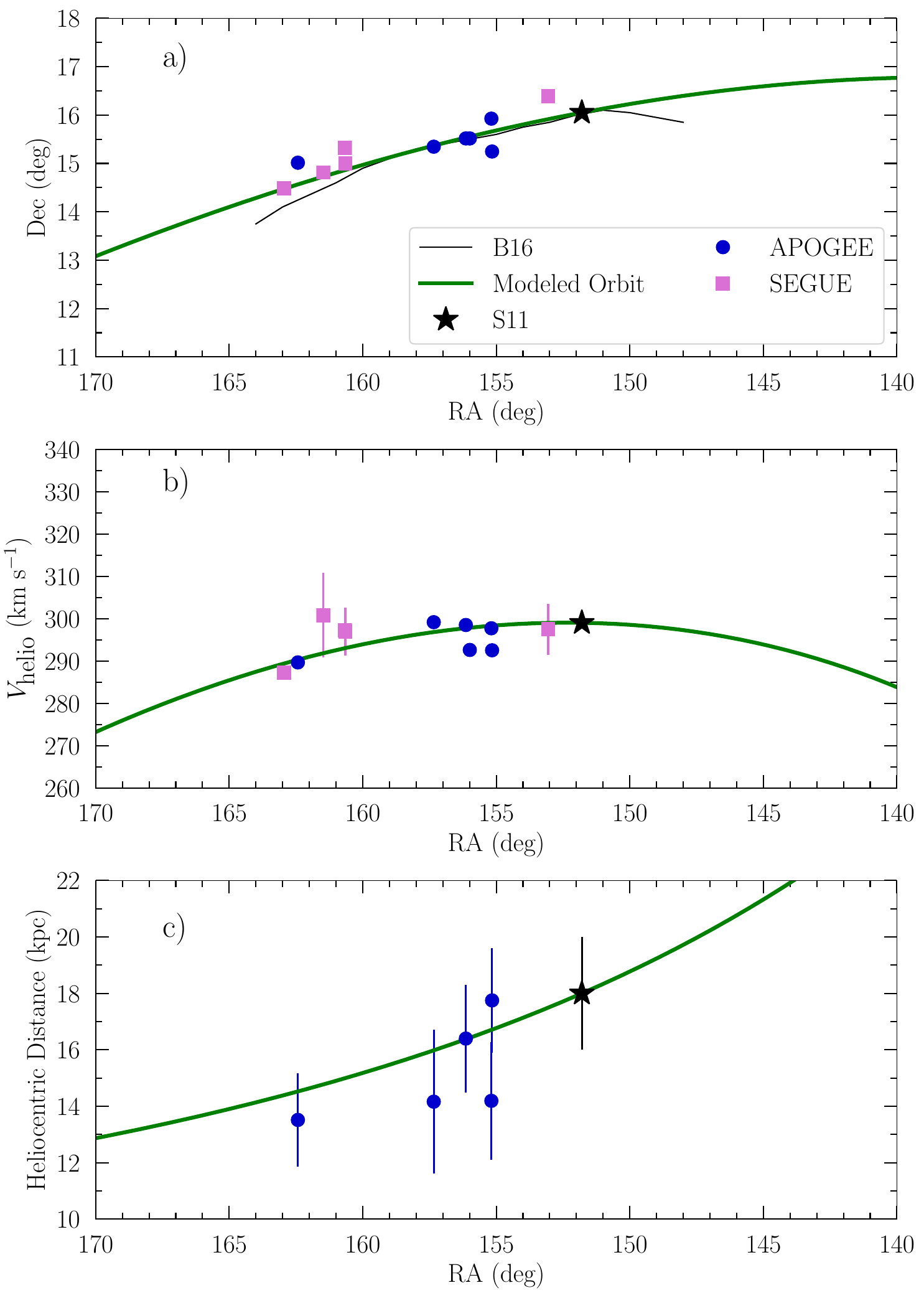}
\caption{(a) Modeled orbit of the stream compared to its observed position on the sky. The orbit passes through the position of the known members, but deviates slightly from the trace of the stream as seen in B16. (b) Modeled heliocentric velocity along the stream track. The modeled orbit is in good agreement with the data. (c) Modeled distance along the stream. Distances plotted are to the APOGEE-2 stream members according to \citet{queiroz2018starhorse}. The modeled orbit is consistent with all of the distance measurements to within $2\sigma$.}
\label{fig:galpyfits}
\end{figure}

\par The panels in Figure~\ref{fig:galpyfits} compare the modeled orbit with the observed properties of the stream. In Figure~\ref{fig:galpyfits}a, the position of the modeled orbit is consistent with the location of the 300S members. The orbit also does not pass close to the position of 2M11081786+1018043, corroborating our decision in Section~\ref{sec:apogee_members} that the star is not a likely 300S member. In Figure~\ref{fig:galpyfits}b and \ref{fig:galpyfits}c, the modeled orbit is consistent with the heliocentric velocity and distance along the stream track. Although the orbit deviates slightly from the trace of the stream as seen in B16, it is still reasonable to use it to infer general features of the stream's kinematic history.

\begin{figure}[!ht] 
\includegraphics[scale=0.4]{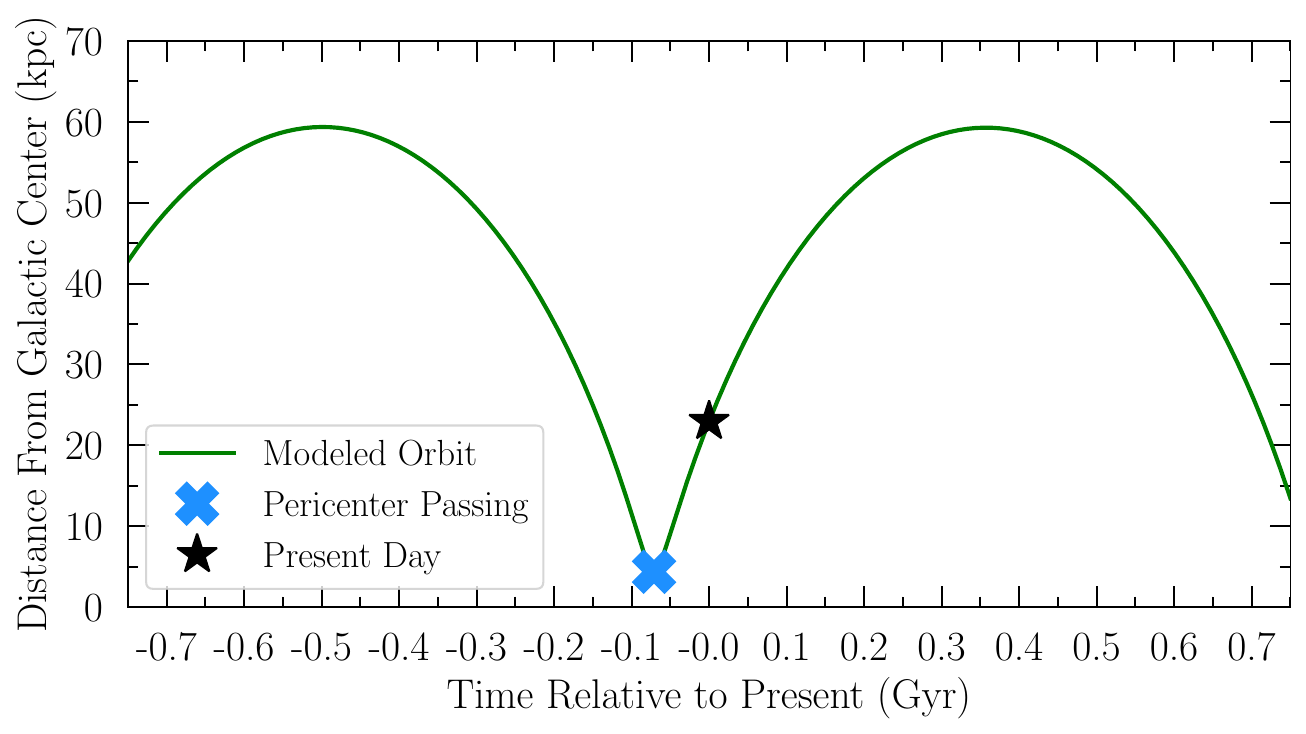}
\caption{Galactocentric distance of the stream orbit as a function of time. The stream passed perigalacticon about 70~Myr ago, approaching within $\sim$4.1~kpc of the Galactic Center.}
\label{fig:time_dist}
\end{figure}

\par Figure \ref{fig:time_dist} shows the distance of the stream away from the Galactic Center as a function of time, suggesting that the progenitor of 300S passed perigalacticon only $\sim$~70~Myr ago. At its closest approach, the progenitor was $\sim$4.1~kpc from the Galactic Center, and thus must have experienced powerful tidal forces. 


\begin{figure*}
\epsscale{1.2}
\plotone{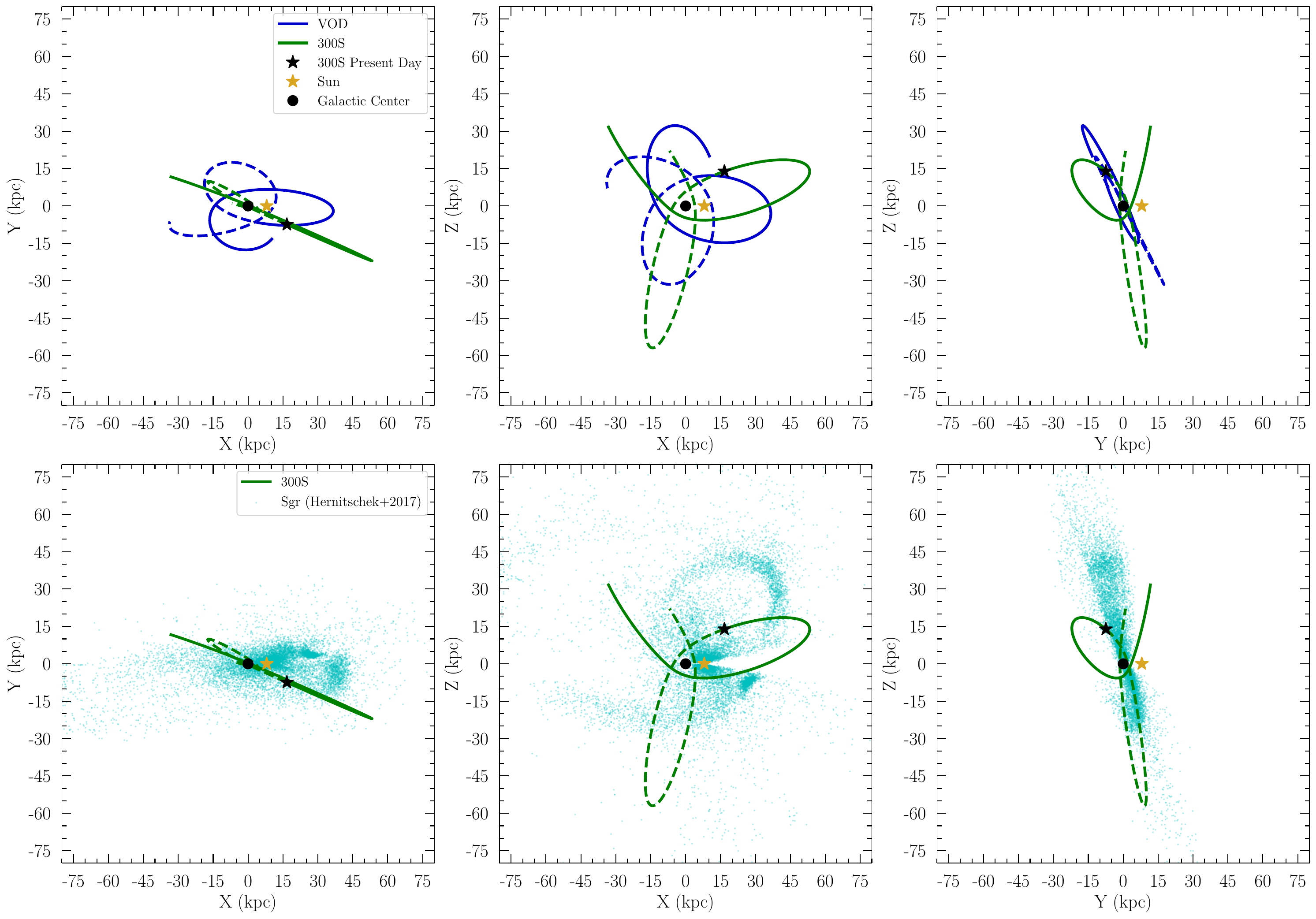}
\caption{Modeled orbit of 300S (green curve) projected onto various Cartesian frames centered at the Galactic Center, illustrating the high eccentricity of the stream's orbit. The top row compares the orbit of 300S with the Virgo Overdensity (VOD), using the data from \citet{carlin2012vod}. The results of this orbit suggest a possible association between 300S and VOD. The second row compares the orbit of 300S with the observed portions of Sagittarius as seen in \citet{sgrobserved}. These results affirm the conclusion of previous studies that Sgr and 300S are unlikely to be kinematically related. More detailed data and modeling would be needed to further ascertain the relationship of 300S to these two substructures.}
\label{fig:galactocentric}
\end{figure*}

\par Figure \ref{fig:galactocentric} shows the modeled orbit projected onto various Cartesian planes centered on the Galactic Center, compared to the orbit of the Sagittarius dwarf spheroidal galaxy (bottom row, observed data from \citep{sgrobserved}) and to that of the Virgo Overdensity \citep[VOD; observed parameters from][]{carlin2012vod}. The orbit of 300S does not resemble that of Sgr \citep[also see][]{law2010sagittarius}, supporting previous suggestions that 300S is unlikely to be kinematically associated with the Sagittarius stream (G09, S11, B16). If the two are related, the 300S progenitor must have been stripped from Sgr long ago. However, the orbit of VOD appears to pass close to 300S, supporting \citet{carlin2012vod}'s suggestion that the two substructures could share a common origin. Finally, the orbit of the stream is perpendicular to the proper motion of Segue~1 \citep{fritz2017orbit}, ruling out any association with that galaxy.

\subsection{Integrated Luminosity of the Stream}

\par In order to place a lower limit on the original mass of the core of the progenitor, we calculate the integrated luminosity of the stream using PS1 photometry \citep{chambers2016pan}. We follow the same procedures outlined in B16 to select stellar-like objects. In particular, we select sources with $| r_{\rm psf} - r_{\rm aperture}| \leq 0.2$~mag. To ensure the quality of our photometry, we also reject sources with uncertainties larger than 0.2~mag in $g$, $r$, and $i$ bands. We correct for reddening effects using the dust maps of \citet{sfd1998} and the extinction law of \citet{schlafly2011}.

\begin{figure}
\epsscale{1.2}
\plotone{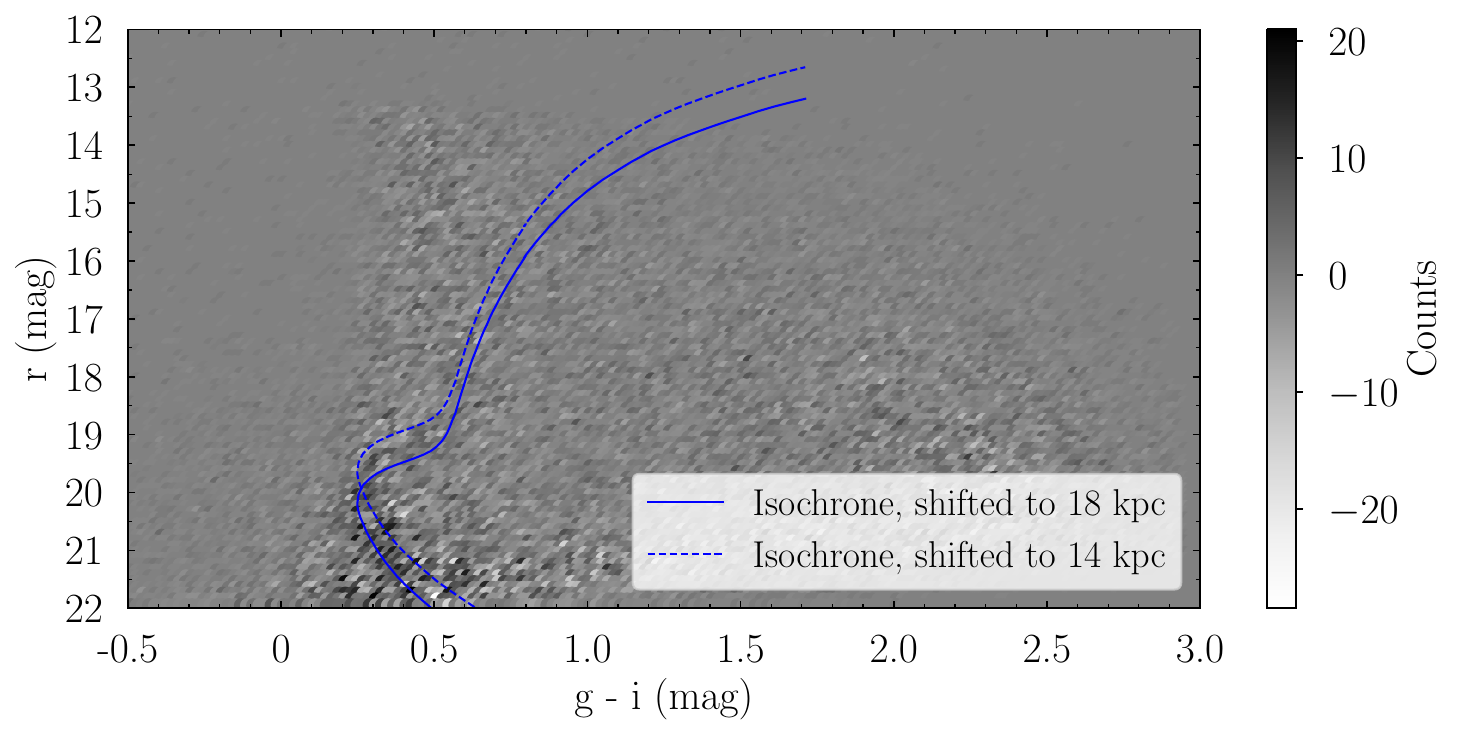}
\caption{Background-subtracted Hess diagram of 300S, where the main sequence of the stream is apparent. The isochrone overplotted at different distance is the same as that in Figure \ref{fig:memdistr}a.}
\label{fig:hess}
\end{figure}

\par We create a Hess diagram of 300S by subtracting a matching Hess diagram of the Milky Way foreground population from the Hess diagram of the region within the stream. The width of the stream is $\sim1\degr$ (see Section~\ref{sec:origin}).  Thus, to construct our foreground distribution, we consider two regions. At a distance of 1\degr\ north of the center of the stream, we obtain the foreground distribution by using the stars within a region of sky that is 1\degr\ wide and runs parallel to the stream track. At a distance of 1\degr\ south of the stream center, we construct another foreground distribution in an analogous fashion. We average the distributions from the regions above and below the stream to produce the foreground Hess diagram, and subtract that from the Hess diagram of the region within the stream. We present the 300S Hess diagram in Figure \ref{fig:hess}, along with the isochrone of 300S overplotted at 14 and 18~kpc for comparison.

\par From the foreground-subtracted Hess diagram, we find a total of 637 stars within the stream down to a magnitude of $r=22$.  For comparison, \citet{martin08} measured 65 stars in Segue~1 ($L = 335$~L$_{\odot}$) down to the same magnitude limit in SDSS.  In each Hess diagram bin along the stream CMD sequence we calculate a luminosity using the isochrone described in Section~\ref{sec:apogee_members} and shown in Figure \ref{fig:memdistr}a.  After correcting for the contribution of sources below the PS1 magnitude limit following \citet{martin08}, we obtain an integrated luminosity for 300S of $4 \times 10^3$~L$_{\odot}$. 

\par Since the chemical abundances of the stream stars suggest that the progenitor of 300S is a dwarf galaxy, we can invoke the mass-metallicity relationship for dwarf galaxies \citep{kirby2013universal} to infer a progenitor stellar mass of $10^{6.9 \pm 0.6}$~M$_{\odot}$, which is on the order of a classical dwarf galaxy (e.g., Leo I, Sculptor). The implied luminosity of this stellar mass is far greater than the observed luminosity of the stream. This suggests that the 300S progenitor was either abnormally metal-rich for its luminosity, or that most of its mass was stripped earlier and is not present in the currently-known part of the stream. 

\section{The Origin of 300S}
\label{sec:origin}

Until the discovery of the ultra-faint dwarf galaxies in 2005
\citep{willman05a, willman05b}, globular clusters and dwarf galaxies
exhibited clearly distinct structural properties, with all dwarf
galaxies having radii of more than 100~pc and all globular clusters
having radii less than 30~pc \citep[e.g.,][]{belokurov07}.  Over the
last decade the size distributions of the two populations have begun
to encroach upon one another at faint luminosities ($L \lesssim
10^{4}$~L$_{\odot}$), such that size alone is no longer sufficient to
classify compact stellar systems.  This convergence has led to the
adoption of alternative classification criteria, namely dynamical
mass-to-light ratio and metallicity dispersion \citep{willman12}.
While the interpretation of the stellar kinematics is rendered more
difficult in the case of a tidally disrupted object, the chemical
properties of the stars are preserved during the disruption process.

To ascertain the nature of the 300S progenitor, we first consider the
chemical abundance measurements presented in Sections~\ref{sec:new_members} and 
\ref{sec:abundances}. The metallicites of the APOGEE-2 member stars are all
quite similar to each other (and to 300S-1 from F13). However, the SEGUE
members span a range in metallicity from $\mbox{[Fe/H]} = -1.3$ to $-2.2$,
such that the intrinsic dispersion of the metallicity distribution for our
full sample of new members is 0.21$_{-0.09}^{+0.12}$~dex.  While we cannot distinguish this value from zero with high statistical significance, the data do suggest that 300S is not a mono-metallic system.  Although the sample size is small, we also do not 
see any sign of the characteristic globular cluster light-element abundance 
correlations in 300S. With detailed abundance patterns for only five stars, 
we cannot exclude the possibility that all of the APOGEE-2 stars happen to belong 
to a single stellar generation and that high-resolution spectroscopy of
additional stars would reveal multiple populations with correlated
abundances. While each individual piece of evidence is not a strong indicator
of the nature of the progenitor, taken together, the best explanation for the 
data is that the progenitor of 300S was a dwarf galaxy. 

To constrain the size of the stream progenitor, we estimate the width
of the stream.  The profile of the stream perpendicular to its length
has a full-width at half maximum of 0.94\degr, corresponding to
a physical extent of $\sim260$~pc at an average distance of 16~kpc.
This size is consistent with the hypothesis that the progenitor was a
relatively compact dwarf galaxy.

S11 measured a velocity dispersion for the stream at the position of
Segue~1 of $7.0 \pm 1.4$~\kms\ with all candidate member stars
included, and $5.6 \pm 1.2$~\kms\ if several possible foreground stars
are excluded.  Because the best-fit stream model we computed in
Section~\ref{sec:new_members} indicates that there may be a velocity gradient of
$\sim10$~\kms\ across the full region spanned by the spectroscopic
data, we cannot simply use the entire sample of member stars to
calculate the velocity dispersion.  Instead, we use the fact that the
five of the six APOGEE-2 stars and four of the five SEGUE stars are
clustered together (near $\alpha_{2000} = 156\degr$ and $\alpha_{2000} =
161.5\degr$, respectively) to determine local velocity dispersions at these
two positions.  The dispersion of the APOGEE-2 stars near
$\alpha_{2000} = 156\degr$ is $\sigma = 3.3^{+1.8}_{-1.1}$~\kms, while that
of the SEGUE stars near $\alpha_{2000} = 161.5\degr$ is $\sigma =
5.1^{+4.8}_{-2.8}$~\kms.  Within the uncertainties, we therefore
conclude that the data are consistent with the stream having a
constant velocity dispersion of $\sim4-5$~\kms\ over the RA range
$\alpha_{2000} = 151.8\degr-161.5\degr$.  Recognizing that the disruption of the
progenitor may have resulted in a stream velocity dispersion that is
higher than that of the progenitor system, we note that this
dispersion is larger than that of the prototypical
globular cluster stream Pal~5 \citep{odenkirchen2009, kuzma2015}.

Finally, we consider the stream orbit derived in Section~\ref{sec:orbit}.  The
path of the stream indicates that it is moving on a highly elliptical
orbit.  Such an orbit would be quite unusual for a globular cluster
\citep[e.g.,][]{dinescu1999,allen2006}.  More eccentric orbits are expected for dwarf
galaxies, which may have fallen into the Milky Way from large
distances, as opposed to forming in situ like many globular clusters.
Orbits approaching within a few kpc of the Galactic Center are likely
not common for dwarfs either, but very small perigalacticon distances
are necessary in order to completely disrupt a dark matter-dominated
system.  As an example, S11 calculated that a galaxy with the mass and
size of Segue~1 would need to pass within $\sim4$~kpc of the Galactic
Center to be disrupted.

We conclude that the observed properties of 300S
favor a dwarf galaxy, rather than a globular cluster, progenitor. Next,
we examine some of its characteristics in the context of other dwarf
galaxies.  If we invoke the mass-metallicity relation for dwarf
galaxies \citep{kirby2013universal}, the metallicity of the stream
corresponds to a progenitor stellar mass of $10^{6.9 \pm 0.6}$~M$_{\odot}$, 
which is comparable to a classical dwarf
spheroidal galaxy such as Leo I (${\rm [Fe/H]} = -1.43$). 
That 300S reaches solar levels of [$\alpha$/Fe] at a metallicity 
between that of Ursa Minor and Sgr also suggests that the stellar 
mass of its progenitor is between $2\times10^5$ M$_{\odot}$ (UMi)  and at least $2\times10^7$ M$_{\odot}$ (Sgr, core). Using Equation 28 from \citet{erkal2016number},
which calculates the mass of the stream progenitor from the width of
the stream on the sky and the enclosed mass at the stream distance,\footnote{The width used in this calculation is the $\sigma$ of a Gaussian fit to the stream profile (D. Erkal 2018, private communication), which we measure to be 0.4\degr.  It is also important to note that \citet{erkal2016number} derived this relation for the case of a single-component (i.e., purely stellar) progenitor.  For a dwarf galaxy progenitor containing both stars and dark matter, the mass determined with this method should correspond to the dynamical mass within a radius comparable to the width of the stream rather than the stellar mass (D. Erkal 2018, private communication).  This value of course may be much smaller than the mass with which the progenitor formed if stripping has been ongoing for a long time.} we obtain a progenitor mass of $10^{5.3}$~M$_{\odot}$.  However,
the integrated luminosity of the stream over its observed extent is
only $4000$~L$_{\odot}$.  Reconciling these numbers requires either
that the progenitor dwarf was unusually metal-rich (and perhaps
unusually extended) for its luminosity, that the progenitor was strongly dark-matter dominated, or that nearly all of the
stars belonging to the progenitor lie outside the known stream.  Since
the stream's orbit extends out to nearly 60~kpc, where most of its
stars would be too faint to be detected by current surveys, the latter
scenario may be plausible.  In addition, it is possible that the
progenitor made previous close passages to the Galactic Center during
which most of its mass was lost.  Given the derived orbital period of
$\sim1$~Gyr (see Fig. \ref{fig:time_dist}), such stars could now be located quite far
away from 300S.

\par The origin of 300S could be connected with other known substructures. While the orbit of 300S is not perfectly coincident with that of VOD, their similarity suggests that that the two substructures could share a common origin. \citet{carlin2012vod} estimated a progenitor mass of $10^{9} M_{\odot}$ for VOD, so 300S may have fallen into the Milky Way with a more massive companion. Cosmological simulations indicate that $\sim30-60$\% of dwarf satellites around MW-like halos were accreted as members of galaxy groups \citep{wetzel2015satellite}; thus, the phenomenon of group infall more generally is not unlikely. Although the radial velocity and orbit shape of 300S are not consistent with the known portions of the Sgr stream (G09), the spatial overlap between the two and their chemical similarity suggests the possibility that the stream progenitor might once have been a dwarf satellite of Sagittarius.  Previous wraps of Sgr debris around the Galaxy are not well-constrained by existing data, so an association with Sgr cannot currently be ruled out observationally. For both VOD and Sgr, improved proper motion measurements and more detailed modeling of the early history of their interaction with the Milky Way would be needed to understand their relationship to 300S.

\section{Conclusions}
\label{sec:conclusion}

\par In this study, we present 11 new members of 300S identified in the APOGEE-2 and SEGUE spectroscopic surveys. From the position of these stars on the sky, we show that 300S is the kinematic counterpart of the elongated photometric substructure found in the same region.

\par We find that the 300S members from APOGEE-2 are chemically similar to Local Group dwarf galaxies and the Milky Way halo, and do not display the characteristic light-element abundance correlations of globular clusters. The new known members also display a metallicity dispersion of 0.21$_{-0.09}^{+0.12}$~dex, exceeding an intrinsic dispersion of 0 by 2 $\sigma$. This suggests that the progenitor may have had an extended period of star formation and a potential well sufficiently deep to retain supernova ejecta. The relatively large width and velocity dispersion of the stream also point to a massive progenitor. Thus, we conclude that 300S is likely the remnant of a tidally disrupted dwarf galaxy. 

\par We infer the proper motion of the stream by fitting the observed properties of the stream to orbits generated from a grid of possible proper motions. The best-fit orbit is highly eccentric, with an apogalacticon distance of 60~kpc and perigalacticon distance of 4.1~kpc away from the Galactic Center. The orbital period of 300S is $\sim1$~Gyr, with its most recent perigalacticon passage 70~Myr ago.

\par Invoking the mass-metallicity relationship for dwarf galaxies, we find that the progenitor of 300S should have a stellar mass of $10^{6.9 \pm 0.6}$~M$_{\odot}$, which is comparable to classical dwarf spheroidal galaxies such as Leo I, Sculptor, and Fornax. We also calculate the integrated luminosity of the stream to be $4 \times 10^3$~L$_{\odot}$, which is much lower than the luminosity implied by the stellar mass from the previous relation. However, at a perigalacticon distance of 4.1~kpc, the tidal field of the Milky Way is sufficiently strong for even a dark matter-dominated system to undergo tidal disruption. With an orbital period of 1~Gyr, it is quite possible that the progenitor of 300S lost most of its stars over multiple close passages to the Milky Way. This is consistent with matched-filter maps of the stream, which show a system that is completely tidally disrupted. 

\par At an observed distance of 20~kpc away from the Milky Way center, 300S may be a valuable probe of the Milky Way potential interior to that distance. With a perigalacticon passage of 4.1~kpc, the orbit of stars in 300S may be affected by time-dependent effects of the Galactic Bar. For reference, the Pal 5 stream, with a perigalacticon distance of of 8~kpc, displays gaps that may have resulted from the bar rotation \citep{pearson2017bar}. Thus, the modeling of 300S in tandem with other stellar streams should provide a more complete picture of the Milky Way potential within its inner tens of kpc. 

\par The upcoming Gaia Data Release 2 should provide strong constraints on the proper motion along the stream track, as the brightest members of 300S are predicted to have proper motion uncertainties of just $\sim0.06$~\masyr\ \citep{gaiadr2}. The Gaia data should also aid in selecting additional 300S targets for spectroscopic followup to determine the velocity dispersion and gradient along the stream, as well as for determining detailed chemical abundances of more stream members. 

\acknowledgements{
S.W.F. thanks Gwen Rudie and Jorge Moreno for their guidance over the course of this work. The authors would like to acknowledge Young Sun Lee for providing [$\alpha$/Fe] and [C/Fe] abundance ratios for the SEGUE stars. They also thank the anonymous referee for comments that improved the specificity and clarity of the paper. They would also like to acknowledge the following individuals for helpful conversations regarding the analysis for this work: Jeff Carlin, Andy McWilliam, Alex Ji, Ana Bonaca, Stephanie Tonnesen, and Rachael Beaton. S.W.F. would also like to acknowledge the maintenance and cleaning staff at academic and telescope facilities, especially those whose communities are often excluded from academic pursuits, but whose labor maintains the spaces where astrophysical inquiry can flourish. 

Funding for S.W.F. was provided by the Pomona Summer Undergraduate Research Program, as well as the Rose Hills Foundation. TCB and VMP acknowledge partial support for this work from grant PHY 14-30152; Physics Frontier Center/JINA Center for the Evolution of the Elements (JINA-CEE), awarded by the US National Science Foundation. JB acknowledges the support of the Natural Sciences and Engineering Research Council of Canada (NSERC), funding reference number RGPIN-2015-05235, and from an Alfred P. Sloan Fellowship. J.G.F-T is supported by FONDECYT No. 3180210. OZ and DAGH acknowledge support provided by the Spanish Ministry of Economy and Competitiveness (MINECO) under grant AYA-2017-88254-P. ARL acknowledges financial support provided by the FONDECYT REGULAR project 1170476

Funding for the Sloan Digital Sky Survey IV has been provided by the Alfred P. Sloan Foundation, the U.S. Department of Energy Office of Science, and the Participating Institutions. SDSS-IV acknowledges support and resources from the Center for High-Performance Computing at the University of Utah. The SDSS web site is www.sdss.org.

SDSS-IV is managed by the Astrophysical Research Consortium for the Participating Institutions of the SDSS Collaboration including the 
Brazilian Participation Group, the Carnegie Institution for Science, 
Carnegie Mellon University, the Chilean Participation Group, the French Participation Group, Harvard-Smithsonian Center for Astrophysics, Instituto de Astrof\'isica de Canarias, The Johns Hopkins University, Kavli Institute for the Physics and Mathematics of the Universe (IPMU) / University of Tokyo, Lawrence Berkeley National Laboratory, Leibniz Institut f\"ur Astrophysik Potsdam (AIP),  Max-Planck-Institut f\"ur Astronomie (MPIA Heidelberg), 
Max-Planck-Institut f\"ur Astrophysik (MPA Garching), Max-Planck-Institut f\"ur Extraterrestrische Physik (MPE), National Astronomical Observatories of China, New Mexico State University, New York University, University of Notre Dame, Observat\'ario Nacional / MCTI, The Ohio State University, Pennsylvania State University, Shanghai Astronomical Observatory, United Kingdom Participation Group, Universidad Nacional Aut\'onoma de M\'exico, University of Arizona, University of Colorado Boulder, University of Oxford, University of Portsmouth, University of Utah, University of Virginia, University of Washington, University of Wisconsin, Vanderbilt University, and Yale University.

The Pan-STARRS1 Surveys (PS1) and the PS1 public science archive have been made possible through contributions by the Institute for Astronomy, the University of Hawaii, the Pan-STARRS Project Office, the Max-Planck Society and its participating institutes, the Max Planck Institute for Astronomy, Heidelberg and the Max Planck Institute for Extraterrestrial Physics, Garching, The Johns Hopkins University, Durham University, the University of Edinburgh, the Queen's University Belfast, the Harvard-Smithsonian Center for Astrophysics, the Las Cumbres Observatory Global Telescope Network Incorporated, the National Central University of Taiwan, the Space Telescope Science Institute, the National Aeronautics and Space Administration under Grant No. NNX08AR22G issued through the Planetary Science Division of the NASA Science Mission Directorate, the National Science Foundation Grant No. AST-1238877, the University of Maryland, Eotvos Lorand University (ELTE), the Los Alamos National Laboratory, and the Gordon and Betty Moore Foundation. The authors wish to recognize and acknowledge the very significant cultural role and reverence that the summit of Maunakea has always had within the indigenous Hawaiian community. We are most fortunate to have the opportunity to utilize observations taken from this mountain. 

This work has made use of data from the European Space Agency (ESA)
mission {\it Gaia} (\url{https://www.cosmos.esa.int/gaia}), processed by
the {\it Gaia} Data Processing and Analysis Consortium (DPAC,
\url{https://www.cosmos.esa.int/web/gaia/dpac/consortium}). Funding
for the DPAC has been provided by national institutions, in particular
the institutions participating in the {\it Gaia} Multilateral Agreement.

This research has made use of NASA's Astrophysics Data System Bibliographic Services and the VizieR catalogue access tool, CDS, Strasbourg, France. The original description of the VizieR service was published in A\&AS 143, 23.
 
This research made use of Astropy, a community-developed core Python package for Astronomy \citep{astropy}.}
 
\bibliography{bibliography}
\bibliographystyle{aasjournal}

\end{document}